\documentclass[reprint,prx,superscriptaddress,twocolumn,showpacs]{revtex4-1}

\usepackage{fullpage} \usepackage{amsmath} \usepackage{amssymb}
\usepackage{xcolor} \usepackage{graphicx} \usepackage{gensymb}
\usepackage{units}
\usepackage{comment}
\usepackage{lipsum}

\usepackage[plainpages=false,pdfpagelabels,colorlinks=true,
linkcolor=red,urlcolor=blue,citecolor=blue,pdftitle={Title},
pdfauthor={},pdfdisplaydoctitle=true,pdfduplex=DuplexFlipLongEdge]{hyperref}

\definecolor{darkred}{rgb}{0.90,0.2,0.2}
\definecolor{darkgreen}{rgb}{0,0.60,.2}
\definecolor{darkblue}{rgb}{0.1,0.3,1}
\definecolor{grey}{cmyk}{0,0,0,0.25}
\definecolor{orange}{cmyk}{0,0.6,0.8,0}

\begin{document}

\title{Emergent eigenstate solution to quantum dynamics far from equilibrium}
\author{Lev Vidmar}
\affiliation{Department of Physics, The Pennsylvania State University, University Park, PA 16802, USA}
\author{Deepak Iyer}
\affiliation{Department of Physics and Astronomy, Bucknell University, Lewisburg, PA 17837, USA}
\author{Marcos Rigol}
\affiliation{Department of Physics, The Pennsylvania State University, University Park, PA 16802, USA}

\begin{abstract}
The quantum dynamics of interacting many-body systems has become a unique venue for the realization of novel states of matter. Here we unveil a new class of nonequilibrium states that are eigenstates of an emergent local Hamiltonian. The latter is explicitly time dependent and, even though it does not commute with the physical Hamiltonian, it behaves as a conserved quantity of the time-evolving system. We discuss two examples of integrable systems in which the emergent eigenstate solution can be applied for an extensive (in system size) time: transport in one-dimensional lattices with initial particle (or spin) imbalance, and sudden expansion of quantum gases in optical lattices. We focus on noninteracting spinless fermions, hard-core bosons, and the Heisenberg model. We show that current-carrying states can be ground states of emergent local Hamiltonians, and that they can exhibit a quasimomentum distribution function that is peaked at nonzero (and tunable) quasimomentum. We also show that time-evolving states can be highly-excited eigenstates of emergent local Hamiltonians, with an entanglement entropy that does not exhibit volume-law scaling.
\end{abstract}

\maketitle

\section{Introduction}

Experiments with ultracold gases \cite{greiner_mandel_02b, kinoshita_wenger_06, will_best_10, trotzky_chen_12, gring_kuhnert_12, will_iyer_15_97, langen_erne_15}, photonic \cite{Christodoulides03, rechtsman_zeuner_13} and solid state systems \cite{wang_steinberg_13, fausti_tobey_11, stojchevska14}, and foundational theoretical developments \cite{polkovnikov_sengupta_review_11, dalessio_kafri_16, eisert_friesdorf_review_15, vidmar16} are driving the study of the dynamics of quantum many-body systems at a rapid pace. When taken far from equilibrium, generic isolated quantum systems typically thermalize \cite{rigol_dunjko_08}, whereas integrable systems (characterized by an extensive number of local conserved quantities) do not. They are instead described by generalized Gibbs ensembles \cite{rigol_dunjko_07, ilievski15}. For nonintegrable systems close to integrable points, one expects relaxation to long-lived states (prethermal states \cite{berges_borsanyi_04}), that can also be described using generalized Gibbs ensembles~\cite{kollar_wolf_11, essler_14, nessi_iucci_14}. Drawing from notions of the equilibrium renormalization group, prethermal states can be understood as being nonthermal fixed points \cite{berges_rothkopt_08}. More recently, the discovery of dynamical phase transitions \cite{heyl_polkovnikov_13, canovi_14, heyl_15}, which are the result of nonanalytic behavior in time, has added another dimension to the connection between quantum dynamics and traditional statistical mechanics.

Here, we add yet another paradigm to this already rich phenomenology. It is motivated by recent theoretical studies that have revealed an intriguing emergence of power-law correlations (like those in ground states) in transport far from equilibrium. Such a phenomenon has been observed in various one-dimensional lattice systems of hard-core \cite{rigol04,rigol05a} and soft-core \cite{rodriguez_manmana_06} bosons, spinful fermions \cite{hm08}, and spins \cite{antal97, lancaster10, sabetta_misguich_13}. Some of the observed behavior can be reproduced using ground states of effective Hamiltonians (different from the ones considered here)~\cite{antal97,antal98,hm08, eisler09, sabetta_misguich_13}. Another motivation for our work are recent experiments exploring the sudden expansion of ultracold fermionic and bosonic gases in optical lattices~\cite{schneider12, ronzheimer13, xia_zundel_15, vidmar15}. One of those experiments, which is of particular relevance to this work, studied the sudden expansion of a Mott insulator of strongly interacting bosons~\cite{vidmar15}. It was observed that peaks develop in the momentum distribution at nonzero momenta, signaling unconventional quasicondensation \cite{rigol04}. Since the expansion in the experiments occurs at energies far above the ground-state energy, it has remained a mystery why quasicondensation (revealing the emergence of power-law correlations) occurs in such systems.

In this work, we provide an explanation for this phenomenon. We unveil the existence of a class of nonequilibrium states that are eigenstates of an emergent local Hamiltonian. We use the term {\it emergent} to highlight that it is not trivially related to the physical Hamiltonian dictating the dynamics. The emergent Hamiltonian is explicitly time dependent and behaves as a local conserved quantity, even though it does not commute with the physical Hamiltonian. The novelty of this class of states is that they exhibit nontrivial time evolution despite being eigenstates of a local conserved operator.

The concept of the emergent eigenstate solution provides new insights into several physical phenomena. It explains why power-law correlations with ground-state character emerge in current-carrying states of integrable (or nearly integrable) one-dimensional models. It also elucidates the dynamics of quasimomentum occupations in current-carrying states, and suggests a way to dynamically tune the position of the peak of the quasimomentum distribution function. In the context of entanglement entropy, the emergent eigenstate solution shifts the focus from the entanglement entropy of time-evolving states to the entanglement entropy of eigenstates of local Hamiltonians. It also highlights the physical relevance of highly-excited eigenstates of integrable systems in which the entanglement entropy does not exhibit a volume-law scaling. 

The paper is organized as follows. In Sec.~\ref{sec_formal}, we introduce the framework of the emergent eigenstate description, and discuss properties of the emergent local Hamiltonian. We also consider conditions for the emergent eigenstate description to be valid. We then discuss two physical applications in Secs.~\ref{sec_transport} and~\ref{sec_suddenexp}. In Sec.~\ref{sec_transport}, we study transport and current-carrying states of noninteracting fermions, hard-core bosons, and in the Heisenberg model. We devote special attention to cases in which the time-evolving state is the ground state of the emergent local Hamiltonian. In Sec.~\ref{sec_suddenexp}, we study the sudden expansion of quantum gases in a setup close to the one realized in recent experiments in optical lattices. Finally, in Sec.~\ref{sec_entanglement}, we focus on the entanglement entropy of the current-carrying states studied in Sec.~\ref{sec_transport}. We show that highly-excited eigenstates that do not exhibit volume-law scaling of the entanglement entropy are the ones of relevance to the problems studied here. We summarize our results, and discuss other possible applications of the emergent eigenstate description, in Sec.~\ref{sec_conclusion}.

\section{Emergent eigenstate description} \label{sec_formal}

\subsection{Construction of the emergent Hamiltonian} \label{construction}

We consider systems initially described by a Hamiltonian
\begin{equation} \label{def_H0}
 \hat H_0 = \hat H + \gamma \hat P \, ,
\end{equation}
where $\hat H$ and $\hat P$ are extensive sums of local operators, namely, of operators with support on $O(1)$ lattice sites. The parameter $\gamma$ may take any value, including $\gamma \to \infty$ (for which $\hat H_0/\gamma \to \hat P$). We therefore do not require any of the two operators $\hat H$ or $\hat P$ to act as a perturbation. We consider initial states $|\psi_0\rangle$ that are eigenstates of $\hat H_0$,
\begin{equation} \label{def_initial}
(\hat H_0 - \lambda)|\psi_0\rangle = 0 \, ,
\end{equation}
where $\lambda$ is the corresponding energy eigenvalue.

In the quantum quenches of interest here, at time $t=0$, the time-independent Hamiltonian $\hat H_0$ is changed instantaneously into the time-independent Hamiltonian $\hat H$ ($\gamma=0$ after the quench), and the initial state evolves as $|\psi(t)\rangle = e^{-i\hat Ht}|\psi_0\rangle$ (we set $\hbar=1$). We relate the operators involved in such quenches by means of the equation
\begin{equation} \label{crit1} 
[\hat H, \hat P] = i a_0 \hat Q \, ,
\end{equation}
where $\hat Q$ is a local operator and $a_0$ is some constant.

We now manipulate Eq.~(\ref{def_initial}), by inserting an identity $\hat I = e^{i\hat H t} e^{-i \hat H t}$ and multiplying by $e^{-i\hat H t}$ on the left, leading to
\begin{equation} \label{defH} 
\left( e^{-i\hat H t} \hat H_0 e^{i\hat H t} - \lambda \right) |\psi(t)\rangle \equiv {\cal \hat M} (t) |\psi(t)\rangle = 0 \, ,
\end{equation}
where ${\cal \hat M}(t)$ is a time-dependent operator in the Schr\"odinger picture (it is time independent in the Heisenberg picture,  ${\cal \hat M}_{\rm H}(t) = \hat H_0 - \lambda$). In general, ${\cal \hat M}(t)$ is highly nonlocal and, hence, of no particular interest. Its nonlocal character is apparent in the series expansion of $e^{-i\hat H t} \hat H_0 e^{i\hat H t}$, which is an infinite sum of nested commutators of $\hat H$ and $ \hat H_0$. Each nonvanishing higher-order commutator usually extends the spatial support of the products of operators involved in ${\cal \hat M}(t)$. Using Eq.~(\ref{crit1}), we can write
\begin{equation} \label{def_M}
 {\cal \hat M}(t) = \hat H_0 - \lambda + \gamma a_0 t\, \hat Q + \gamma a_0 \sum_{n=1}^\infty (-i)^n \frac{t^{n+1}}{(n+1)!} {\cal \hat H}_n,
\end{equation}
where ${\cal \hat H}_n$ represents the $n$-th order commutator of $\hat H$ with $\hat Q$
\begin{equation} \label{HQcommutators}
{\cal \hat H}_n  = \underbrace{[\hat H,[\hat H, \ldots [\hat H, \hat Q] \ldots ]]}_{\text{$n$ commutators}}.
\end{equation} 

Even though the expansion~(\ref{def_M}) has, in general, zero radius of convergence, there are physically relevant problems for which ${\cal \hat M}(t)$ becomes a local operator. This can occur either if ${\cal \hat H}_n$ vanishes at some finite $n_0$ or if the nested commutators close the sum~(\ref{def_M}). If ${\cal \hat M}(t)$ is a local operator, we define ${\cal \hat H}(t)\equiv{\cal \hat M}(t)$. Physically, we interpret ${\cal \hat H}(t)$ as being an {\it emergent local Hamiltonian}. ${\cal \hat H}(t)$ is a local {\it conserved} operator of the time-evolving system (it is time independent in the Heisenberg picture, ${\cal \hat H}_{\rm H}(t) = \hat H_0 - \lambda$). Its novelty comes from the fact that, despite being conserved, it does not commute with the Hamiltonian $\hat H$ that governs the dynamics. This is possible only because ${\cal \hat H}(t)$ is time dependent in the Schr\"odinger picture.

Whenever the emergent Hamiltonian description can be invoked, instead of explicitly time evolving the wavefunction, one only needs to find a single eigenstate $|\Psi_t \rangle$ of ${\cal \hat H}(t)$ satisfying
\begin{equation} \label{HemePsi}
{\cal \hat H}(t) |\Psi_t \rangle = 0 \, .
\end{equation}
Equations~(\ref{def_H0})-(\ref{defH}) imply that, in the absence of degeneracies, this eigenstate is $|\psi(t)\rangle$. We call this scenario the {\it emergent eigenstate solution to quantum dynamics}. It gives rise to a new class of nonequilibrium states: states that simultaneously exhibit nontrivial time evolution and are eigenstates of a time-dependent local operator that is conserved during the time evolution of the system. By conserved we mean that the expectation value of ${\cal \hat H} (t)$ is time independent under dynamics generated by $\hat H$.

The simplest family of quantum quenches in which the emergent eigenstate solution to quantum dynamics can be used is the one in which
\begin{equation}
{\cal \hat H}_1 = [\hat H, \hat Q] = 0 \, .
\end{equation}
It results in an emergent local Hamiltonian of the form
\begin{equation} \label{Heme} 
{\cal \hat H} (t) =  \hat H_0 -\lambda + \gamma a_0 t \; \hat Q  \, .
\end{equation}

There are families of quantum quenches for which $\hat Q$ is not exactly conserved (${\cal \hat H}_n$ is nonzero for all $n$), but for which the time-dependent state is exponentially close to an eigenstate of the emergent local Hamiltonian for times that are proportional to the system size. In the context of physical applications, we  are interested in families of quantum quenches that fall into this category. In such quenches $\hat Q$ is conserved up to boundary terms. Next, we discuss a criterion for the applicability of the emergent eigenstate description for those quenches, and clarify the role of the initial state.

\subsection{Emergent eigenstate description for approximately conserved operators} \label{sec_almost}

If $\hat Q$ is not exactly conserved, i.e., ${\cal \hat H}_n \neq 0$ for all $n$, but it is approximately conserved (e.g., $[\hat H, \hat Q]$ results in terms with support only at the boundaries of the system), one may still invoke the emergent eigenstate description by truncating the series~(\ref{def_M}) at some $n_0$. For the discussion in this section, we assume that one can truncate the series at $n_0=1$. As a result, the emergent local Hamiltonian ${\cal \hat H}(t)$ has the form~(\ref{Heme}). In principle, such ${\cal \hat H}(t)$ is not conserved since in the Heisenberg picture
\begin{equation} \label{def_Hheis_general}
 {\cal \hat H}_{\rm H}(t) = \hat H_0 - \lambda + \gamma a_0 \sum_{n=1}^\infty i^n \frac{n \, t^{n+1}}{(n+1)!} {\cal \hat H}_n \, .
\end{equation}

Nevertheless, there are families of initial states for which ${\cal \hat H}(t)$ behaves as being approximately conserved for an extensive (in system size) time. In these cases, one can show that
\begin{equation} \label{Hheis_almost}
 \langle \psi(t) | {\cal \hat H}(t) | \psi(t) \rangle = \varepsilon(t),
\end{equation}
where
\begin{equation} \label{Hheis_almost1}
 \lim_{L\rightarrow\infty}\varepsilon(t)=0\ \  \forall\ \ t<\infty
\end{equation}
and $L$ denotes the (linear) system size.

To see how Eqs.~(\ref{Hheis_almost}) and (\ref{Hheis_almost1}) imply that eigenstate $|\Psi_t\rangle$ of the emergent local Hamiltonian ${\cal \hat H}(t)$ becomes indistinguishable from the time-evolving state $|\psi(t) \rangle$ with increasing system size, let us write $|\psi(t)\rangle$ in the basis defined by the eigenstates of ${\cal \hat H}(t)$:
\begin{equation} \label{psi_Heme_basis}
 |\psi(t) \rangle = \sqrt{1-\eta_t} \, |\Psi_t\rangle + \sum_{m=2}^{\cal D} c_t^{(m)} |\Psi_t^{(m)} \rangle,
\end{equation}
where, from the normalization of $|\psi(t) \rangle$, we have that $\eta_t = \sum_{m=2}^{\cal D} |c_t^{(m)}|^2$, $\eta_t \in [0,1]$, and $\cal D$ is the corresponding Hilbert-space dimension.

Using Eqs.~\eqref{HemePsi} and~\eqref{psi_Heme_basis}, Eq.~(\ref{Hheis_almost}) can be rewritten as
\begin{equation} \label{app_eq2}
 \langle \psi(t) | {\cal \hat H}(t) |\psi(t) \rangle = \sum_{m=2}^{\cal D}  \left|c_t^{(m)}\right|^2 E_t^{(m)} =\varepsilon(t),
\end{equation}
where $E_t^{(m)}$ is the eigenenergy corresponding to eigenstate $|\Psi_t^{(m)} \rangle$, ${\cal \hat H}(t) |\Psi_t^{(m)} \rangle = E_t^{(m)} |\Psi_t^{(m)} \rangle$. We then see that for Eq.~(\ref{Hheis_almost1}) to be satisfied one generally needs 
\begin{equation} \label{def_coefs_Heme}
 \lim_{L\rightarrow\infty} \sum_{m=2}^{\cal D}  \left|c_t^{(m)}\right|^2 =0 \, ,
\end{equation}
yielding $\lim_{L\rightarrow\infty}  \eta_t = 0$.

Equation~(\ref{Hheis_almost}) can also be written as the expectation value of ${\cal \hat H}_{\rm H}(t)$ in the initial state as
\begin{equation} \label{Hheis_initial}
 \langle \psi_0 | {\cal \hat H}_\text{H}(t) | \psi_0 \rangle = \gamma a_0 \sum_{n=1}^\infty i^n \frac{n \, t^{n+1}}{(n+1)!} \langle \psi_0 | {\cal \hat H}_n | \psi_0 \rangle = \varepsilon(t) \, .
\end{equation}
Hence, even though ${\cal \hat H}_{\rm H}(t)$ is explicitly time dependent, the emergent local Hamiltonian can behave as a conserved quantity in some nonequilibrium states for extensively (in the system size) long times.

For most practical applications, the emergent eigenstate description provides a useful framework if the lowest-order terms in Eq.~(\ref{Hheis_initial}) are exactly zero. Moreover, an extensive time of validity will be obtained if an extensive number of expectation values $\langle \psi_0 | {\cal \hat H}_n | \psi_0 \rangle$ vanish independently. This can be realized for systems in which $\hat Q$ is conserved up to boundary terms (see the next two sections).

The concept of the emergent eigenstate description formulated in this section opens a new window for studies of dynamics far from equilibrium. For example, if $|\psi(t)\rangle$ were the ground state of an emergent local Hamiltonian, then all the tools developed to study ground states of quantum systems could immediately be applied to understand a system far from equilibrium. As we discuss next, there are experimentally relevant time-evolving states that are ground states of such emergent local Hamiltonians.

In the following, we report two applications of the emergent eigenstate solution: we study transport and current-carrying states in Sec.~\ref{sec_transport}, and the sudden expansion of quantum gases in optical lattices in Sec.~\ref{sec_suddenexp}. In both applications, we make use of the property that even though $\hat Q$ is not exactly conserved (because of boundary terms), the emergent Hamiltonian satisfies Eq.~(\ref{Hheis_almost}) with an exponentially small $\varepsilon(t)$ over extensively (in the system size) long times. In Sec.~\ref{sec_entanglement}, we study entanglement properties of the emergent local Hamiltonian.

\section{Transport in integrable lattice systems} \label{sec_transport}

Transport of particles (or spin, energy, etc) far from equilibrium is a topic that has been attracting increasing attention in the context of isolated quantum systems. A convenient way to simulate current-carrying states in such systems is to prepare initial states with particle (spin, energy) imbalance that mimics large reservoirs, and let them evolve under homogeneous Hamiltonians. One of the most studied setups is the ``melting'' dynamics of a sharp domain wall~\cite{antal99, karevski02, ogata02, rigol04, hunyadi04, hm08, eisler13, mossel10, santos11, vasseur15, hauschild15}. This setup is a particular case of the more general one considered below.

\subsection{Noninteracting spinless fermions and hard-core bosons} \label{sec_free}

We first construct an emergent eigenstate solution for systems of noninteracting spinless fermions and hard-core bosons. We focus on one-dimensional chains with $L=2N+1$ sites ($N$ is the particle number) and open boundary conditions. We prepare the initial state with particle imbalance by applying a linear gradient along the chain~\cite{eisler09, lancaster10,lancaster16}.

The initial state for noninteracting spinless fermions (SF) is the ground state of
\begin{equation} \label{def_H0_sf}
\hat H_{0,\rm SF} = \hat H_{\rm SF} + \gamma \hat P_{\rm SF},
\end{equation}
where the kinetic energy and the potential energy terms, respectively, are
\begin{eqnarray}
\hat H_{\rm SF} & = & -J \sum_{l=-N}^{N-1} (\hat f_{l+1}^\dagger \hat f^{}_l +{\rm H.c.}) \, , \label{def_Hsf} \\
\hat P_{\rm SF} & = & \frac{1}{L} \sum_{l=-N}^{N} l \, \hat n_l \, . \label{def_Pfree}
\end{eqnarray}
Here, $\hat f_l^\dagger$ creates a spinless fermion at site $l$, and the site occupation operator is $\hat n_l=\hat f_l^\dagger \hat f^{}_l$. The prefactor $1/L$ in Eq.~\eqref{def_Pfree} ensures that the expectation value of $\hat P_{\rm SF}$ is extensive in system size. The strength of the linear gradient $\gamma$ is measured in units of the hopping amplitude $J$, and we set $J=1$. In the limit $\gamma \to \infty$, we simplify~(\ref{def_H0_sf}) and consider $\hat H_{0,\rm SF} = \hat P_{\rm SF}$, see Appendix~\ref{app0} for details.

We require $\gamma$ to be large enough such that, in the ground state of $\hat H_{0,\rm SF}$, there exist regions with site occupations one and zero at the chain boundaries. This is achieved for $\gamma > \gamma^* = 4$, where $\gamma^*$ is the critical value needed for the $N$th single-particle Wannier-Stark state to be a Bessel function with support on $L=2N+1$ sites. The time evolution of such initial state under $\hat H_{\rm SF}$ produces a current-carrying state $|\psi(t)\rangle$. We are interested in this current-carrying state before the propagating front of particles reaches the chain boundary.

We also study the same setup for hard-core bosons, which can be mapped onto spin-1/2 systems, and noninteracting spinless fermions~\cite{cazalilla_citro_review_11}. The interest in this model, which is the infinite on-site repulsion limit of the Bose-Hubbard model, is two fold: (i) it has been studied in experiments with ultracold atoms~\cite{ronzheimer13,vidmar15}, and (ii) in and out of equilibrium, the correlation functions can be very different from those of noninteracting fermions \cite{cazalilla_citro_review_11, vidmar16}. 

For hard-core bosons, we replace $\hat H_{\rm SF}$~(\ref{def_Hsf}) by
\begin{equation}
\hat H_{\rm HCB} = - \sum_{l=-N}^{N-1} (\hat b_{l+1}^\dagger \hat b^{}_l + {\rm H.c.}),
\end{equation}
while, as a consequence of the mapping, the potential term is identical to ${\hat P}_{\rm SF}$~(\ref{def_Pfree}). Infinite repulsion is enforced by the constraints $(\hat b^{}_{l})^2 = (\hat b_{l}^\dagger)^2 = 0$, where $\hat b_{l}^\dagger$ is the boson creation operator at site $l$. We calculate expectation values of observables by expressing operators in terms of spinless fermions and following Refs.~\cite{rigol04,rigol05a}.

Within the setup in this section, Eq.~\eqref{crit1} results in $a_0 =-1/L$ and
\begin{equation} \label{def_Qsf}
\hat Q_{\rm SF} = \sum_{l=-N}^{N-1} (i\hat f_{l+1}^\dagger \hat f^{}_l + {\rm H.c.}) \, ,
\end{equation}
which is the particle current operator in an open lattice. In finite systems with open boundaries, $\hat Q_{\rm SF}$ is not exactly conserved since
\begin{equation} \label{HQfree}
{\cal \hat H}_{1,\rm SF} = [\hat H_{\rm SF},\hat Q_{\rm SF}] = -2i (\hat n_{-N} - \hat n_N) \, .
\end{equation}
As a result, all higher-order commutators ${\cal \hat H}_{n,\rm SF}$~(\ref{HQcommutators}) are nonzero. However, we show in the following that an extensive (in system size) number of expectation values of ${\cal \hat H}_{n>1,\rm SF}$ vanish in the initial state. Equation~(\ref{def_M}) can then be truncated at $n=2$ to get the emergent local Hamiltonian
\begin{eqnarray} \label{emergentH} 
&&{\cal \hat H}_{\rm SF}(t) = - \sum_{l=-N}^{N-1} (\hat f_{l+1}^\dagger \hat f^{}_l + {\rm H.c.}) \, - \lambda 
+ \, \frac{\gamma}{L} \left( \sum_{l=-N}^{N} l \, \hat n_l \,\right.\nonumber\\&&\quad\ -\left. t \sum_{l=-N}^{N-1} (i\hat f_{l+1}^\dagger \hat f^{}_l + {\rm H.c.} ) + t^2 (\hat n_{-N} - \hat n_N) \right).
\end{eqnarray}

Note that ${\cal \hat H}_{\rm SF}(t)$ includes the difference between site-occupation operators at the lattice boundaries (times $t^2$). As we argue below, the emergent eigenstate solution is accurate as long as the propagating front of particles (holes) does not reach the chain boundary, i.e., as long as $\langle \hat n_{-N}(t) \rangle = 1$ and $\langle \hat n_{N}(t) \rangle = 0$. Therefore, in what follows, we replace $\hat n_{-N} \to 1$ and $\hat n_{-N}\to 0$ in Eq.~(\ref{emergentH}), which ensures that the target eigenstate $|\Psi_t\rangle$ of the ensuing emergent local Hamiltonian ${\cal \hat H}_{\rm SF}'(t) $,
\begin{eqnarray}
{\cal \hat H}_{\rm SF}'(t) & = & - {\cal A}(t) \sum_{l=-N}^{N-1} (e^{i \varphi(t)} \hat f_{l+1}^\dagger \hat f^{}_l + {\rm H.c.}) \, \nonumber \\
&  & + \, \frac{\gamma}{L} \sum_{l=-N}^{N} l \, \hat n_l - \left( \lambda - \frac{\gamma t^2}{L} \right) \, ,  \label{Heme_free}
\end{eqnarray}
is the ground state. $|\Psi_t \rangle$ is the ground state of $\hat {\cal H}'_{\rm SF}(t)$ because the latter state is nondegenerate at all times. Hence, since $|\Psi_t\rangle$ is the ground state at $t=0$, it must be the ground state at all times.

In Eq.~(\ref{Heme_free}), we merged the kinetic and the current operators from Eq.~\eqref{emergentH} into a single operator, characterized by the hopping amplitude
\begin{equation}
{\cal A}(t) = \sqrt{1 + (\gamma t/L)^2} \label{def_At}
\end{equation}
and the phase
\begin{equation} \label{def_phase}
\varphi(t) = \arctan{\left(\frac{\gamma t}{L}\right)}.
\end{equation}
In the ground state of ${\cal \hat H}_{\rm SF}'(t) $, the phase $\varphi(t)$ determines the position of the maximum of the quasimomentum distribution function. 

To prove that ${\cal \hat H}_{\rm SF}(t)$ and ${\cal \hat H}_{\rm SF}'(t)$ are indeed relevant to the dynamics of interest here, we computed the expectation values of higher-order commutators ${\cal \hat H}_{n,\rm SF}$ in the initial state~(\ref{Hheis_initial}). The results are reported in Appendix~\ref{app1} for the case in which $\gamma^{-1}=0$, i.e., for the initial sharp domain wall. The analysis reveals that the emergent eigenstate description is exponentially accurate for $t \lesssim N(2/e)$. Since the maximal group velocity in the lattice is $v_{\text{max}}=2$ (in units of $Ja/\hbar$, where $a$ is the lattice spacing), the physical picture consistent with this time restriction is that the emergent Hamiltonian description~(\ref{emergentH}) is valid so long as the expanding particles (holes) do not reach the edge of the lattice, $t\lesssim N/v_\text{max}=N/2$. In that case, the dynamics is expected to be identical to that of a semi-infinite domain wall, for which the boundaries are irrelevant.

\begin{figure}[!t]
  \begin{center}
    \includegraphics[width=1.00\columnwidth]{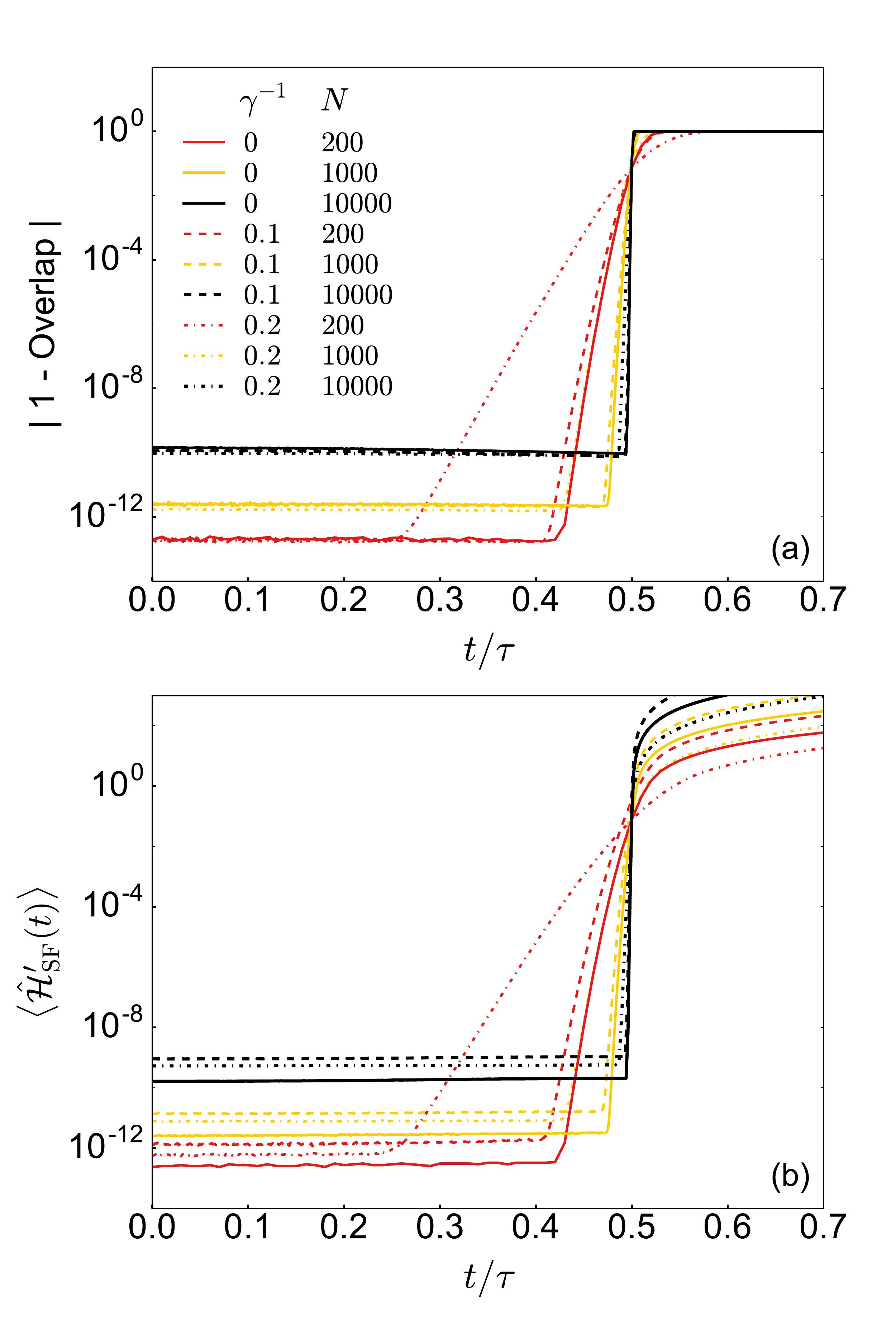}
  \vspace{-0.5cm}
\caption{{\it Times of validity of the emergent eigenstate description.}
(a) Subtracted overlap $|1-O(t)|$, where  $O(t) = |\langle \Psi_t | \psi(t) \rangle |$, of the time-evolving state $|\psi(t)\rangle$ with the ground state $|\Psi_t\rangle$ of the emergent local Hamiltonian ${\cal \hat H}_{\rm SF}'(t)$~(\ref{Heme_free}). (b) Expectation value of the emergent local Hamiltonian ${\cal \hat H}_{\rm SF}'(t)$ in the time-evolving state $|\psi(t)\rangle$. Results in both panels are shown for different values of $\gamma^{-1}=0, 0.1,$ and $0.2$, and three different system sizes $L = 2N+1$, as indicated in the legend. For $\gamma^{-1}=0$, ${\cal \hat H}_{\rm SF}'(t)$ is taken as explained in Appendix~\ref{app0}. We rescale time dividing by $\tau$~(\ref{def_tau}), which is proportional to the system size, to demonstrate the validity of the emergent eigenstate solution for $t/\tau \lesssim 0.5$  when $L \to \infty$.
  \label{fig1} }
   \end{center}
\end{figure}

We complement the analytical results in Appendix~\ref{app1}, and the physical picture that has emerged from them, by numerically calculating the subtracted overlap $1-O(t) =1- |\langle \Psi_t | \psi(t) \rangle |$ and the expectation value $\langle \psi(t)| {\cal \hat H}_{\rm SF}'(t) | \psi(t) \rangle$. In Fig.~\ref{fig1}, we plot the numerical results for those quantities versus the rescaled time $t/\tau$ for various system sizes, where $\tau$ is given by the expression \cite{eisler09,lancaster10} 
\begin{equation} \label{def_tau}
 \tau = N \sqrt{1- \left( \frac{\gamma^*}{\gamma} \right)^2}.
\end{equation}
Note that $\tau = N$ for $\gamma^{-1}=0$. 

In Fig.~\ref{fig1}, $1-O(t)$ and $\langle \psi(t)| {\cal \hat H}_{\rm SF}'(t) | \psi(t) \rangle$ are zero within machine precision at short times and start departing from zero when $t/\tau$ approaches $1/2$, as advanced by our physical picture for $\gamma^{-1}=0$. The same argument applies for nonzero $\gamma^{-1}$. However, that case is more involved since: (i) the initial state exhibits a metallic interface between the left/right regions with maximal/vanishing site occupations, and (ii) the velocity of the propagating front is not constant [see Fig.~\ref{figsup1}(a) in Appendix~\ref{app1b}]. In Appendix~\ref{app1b}, we show that the time $t^*$ at which the site occupations at the boundaries of the lattice depart from their initial values is $t^*/\tau = 1/2$.

The results in Fig.~\ref{fig1} demonstrate that, as long as the emergent local Hamiltonian ${\cal \hat H}_{\rm SF}'(t)$ behaves as a conserved operator, which is the case whenever particles (or holes) have not reached the chain boundary, the ground state of ${\cal \hat H}_{\rm SF}'(t)$ is indistinguishable from the time-evolving state $|\psi(t)\rangle$, as advanced in Sec.~\ref{sec_almost}.

We now turn our focus to physical properties of current-carrying states that can be described using the emergent eigenstate solution. We study one-particle properties such as site occupations, particle currents, decay of one-body correlations, and the quasimomentum distribution function.

\begin{figure}[!t]
  \begin{center}
    \includegraphics[width=1.00\columnwidth]{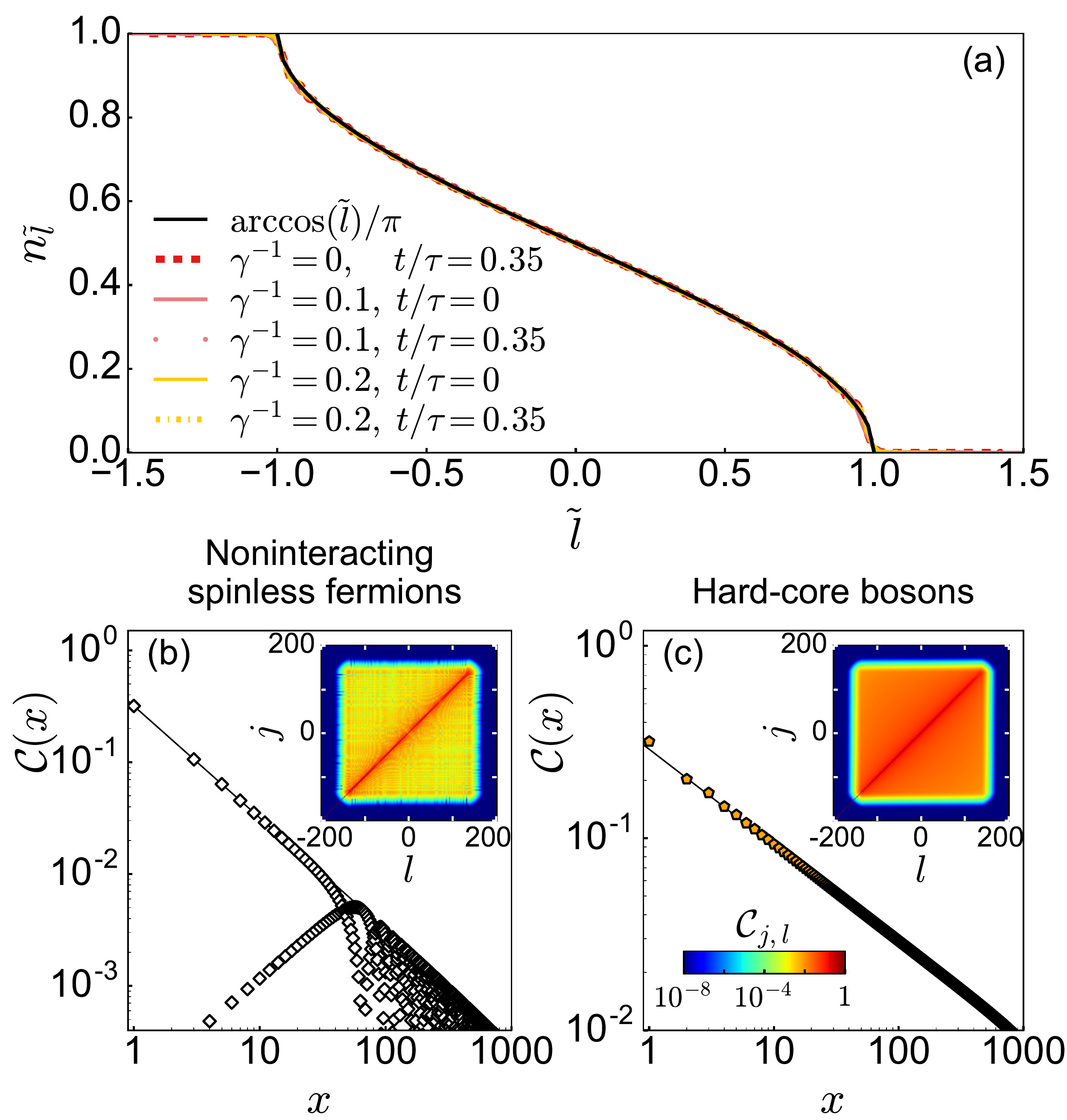}
  \vspace{-0.5cm}
\caption{{\it Site occupations and nonlocal correlations in current-carrying states.}
(a) Site occupations $n_l(t) = \langle \psi(t) | \hat n_l | \psi(t) \rangle$ versus $\tilde l = l \gamma(t)/(2L)$. Results are shown for $N=200$ particles, different values of $\gamma$, and different times. The black solid line is the scaling solution $n_{\tilde l}(t) = \arccos(\tilde l)/\pi$ for $-1<\tilde l < 1$. The main panels in (b) and (c) display ${\cal C}(x) \equiv {\cal C}_{0,x}$, where ${\cal C}_{j,l} = \vert\langle \hat f_{j}^\dagger \hat f^{}_{l}\rangle \vert$ for noninteracting spinless fermions (b) and ${\cal C}_{j,l} = \vert\langle \hat b_{j}^\dagger \hat b^{}_{l}\rangle \vert$ for hard-core bosons (c), as a function of $x$ for $N=2000$ particles, $\gamma^{-1}=0$, at time $t/\tau=0.35$. The solid lines overlapping with the numerical results are ${\cal C}(x)=1/(\pi x)$ in (b) and ${\cal C}(x)=0.29/ \sqrt{x}$ in (c)~\cite{lancaster10}. The insets in (b) and (c) display the absolute values of all elements of the one-body density matrix for the same model parameters and time as the main panels, but calculated for $N=200$ particles.
  \label{fig2} }
   \end{center}
\end{figure}

The mapping of hard-core bosons onto noninteracting spinless fermions~\cite{cazalilla_citro_review_11} implies identical site occupations for bosons and fermions. Their dynamics, $n_l(t) = \langle \psi(t)| \hat n_l |\psi(t) \rangle$, from initial states with particle imbalance has been studied in the past~\cite{antal99,karevski02,ogata02,rigol04,hunyadi04,eisler09,lancaster10}. In many instances, plotting site occupations versus site positions divided by a function of the evolution time results in data collapse. In the setup under consideration here, this is achieved (for any $\gamma > \gamma^*$) by plotting site occupations versus $\tilde l = l \gamma(t)/(2L)$, where $\gamma(t) = \gamma/{\cal A}(t)$. This results in a time- and $\gamma$-independent site occupation profile $n_{\tilde l}(t) = \arccos(\tilde l)/\pi$ for $-1<\tilde l < 1$, as shown in Fig.~\ref{fig2}(a). One can use $n_{\tilde l}(t)$ to derive the velocity of the propagating front of particles (holes) as a function of time [see Appendix~\ref{app1b} and Fig.~\ref{figsup1}(a)].

Another observable that yields identical expectation values for noninteracting fermions and hard-core bosons is the particle current $\hat Q_{\rm SF}$~(\ref{def_Qsf}). Since $\hat Q_{\rm SF}$ does not commute with $\hat H_{\rm SF}$ on a lattice with open boundaries, its expectation value $\langle \psi(t) | \hat Q_{\rm SF} | \psi(t) \rangle$ is time dependent~\cite{antal99}. By invoking the Heisenberg representation and properties of higher-order commutators of $\hat H_{\rm SF}$ with $\hat Q_{\rm SF}$, as discussed in Appendix~\ref{app1}, one gets that
\begin{equation} \label{Q_t_linear}
\langle \psi(t) | \hat Q_{\rm SF} | \psi(t) \rangle = 2 t \, \langle \psi_0 | \left( \hat n_{-N} - \hat n_N \right) |\psi_0 \rangle,
\end{equation}
i.e., the particle current increases linearly in time and is proportional to the difference in site occupations at boundary. This result was generalized in Ref.~\cite{vasseur15} to different families of current-carrying states, including interacting spinless-fermion systems. The relation~(\ref{Q_t_linear}) can already be inferred from Eq.~(\ref{emergentH}) by requiring the expectation value of the emergent local Hamiltonian on the time-evolved state to be time independent. This suggests that Eq.~(\ref{Q_t_linear}) is accurate for the same times for which the emergent eigenstate description is, and it breaks down when the front of propagating particles (holes) reaches the lattice boundary. Hence the particle current in this setup belongs to the class of observables whose expectation values out of equilibrium are controlled, for an extensive (in system size) time, by expectation values of observables in the initial state.

Next we study one-body correlations in the current-carrying states. This is of particular interest since an intimate relation between current-carrying states and power-law correlations has been observed for over the last thirty years~\cite{rigol04, rodriguez_manmana_06, hm08, antal97, lancaster10, sabetta_misguich_13, schmittmann95, spohn83,prosen_znidaric_10}. For noninteracting fermions~\cite{antal99} and hard-core bosons~\cite{rigol04}, previous studies revealed that the emergent (when starting from the sharp domain wall) power-law correlations in the current-carrying states exhibit ground-state exponents (ground state with respect to the physical Hamiltonian). This is an intriguing result because: (i) in thermal equilibrium in one dimension, power-law correlations can only be found in the ground state, and (ii) the current-carrying states considered here have energy densities well above that of the ground state of the physical Hamiltonian. The emergent eigenstate solution introduced in this work explains why ground-state-like correlations emerge during the far-from-equilibrium dynamics: the time-evolving states are ground states of emergent local Hamiltonians.

Figures~\ref{fig2}(b) and~\ref{fig2}(c) show the spatial decay ${\cal C}(x)$ of one-body correlations at time $t/\tau = 0.35$ for a sharp domain wall $\gamma^{-1}=0$ (a product state with no correlations at all) as initial state. We define ${\cal C}(x) \equiv {\cal C}_{0,x}$, where ${\cal C}_{0,x} = \vert\langle \hat f_{0}^\dagger \hat f^{}_{x}\rangle \vert$ for noninteracting fermions and ${\cal C}_{0,x} = \vert\langle \hat b_{0}^\dagger \hat b^{}_{x}\rangle \vert$ for hard-core bosons. For the former [Fig.~\ref{fig2}(b)], one-body correlations decay as ${\cal C}(x)=|\sin(\pi x/2)|/(\pi x)$~\cite{antal99}, while for the latter [Fig.~\ref{fig2}(c)], they decay as ${\cal C}(x)=0.29/\sqrt{x}$~\cite{lancaster10}. The spatial decay of ${\cal C}(x)$ (for $0<n_l<1$) is identical for all $\gamma > \gamma^*$ and $t/\tau < 0.5$. The insets in Figs.~\ref{fig2}(b) and~\ref{fig2}(c) display density plots of the absolute values of all elements of the one-body density matrix.

\begin{figure*}[!]
  \begin{center}
  \includegraphics[width=1.80\columnwidth]{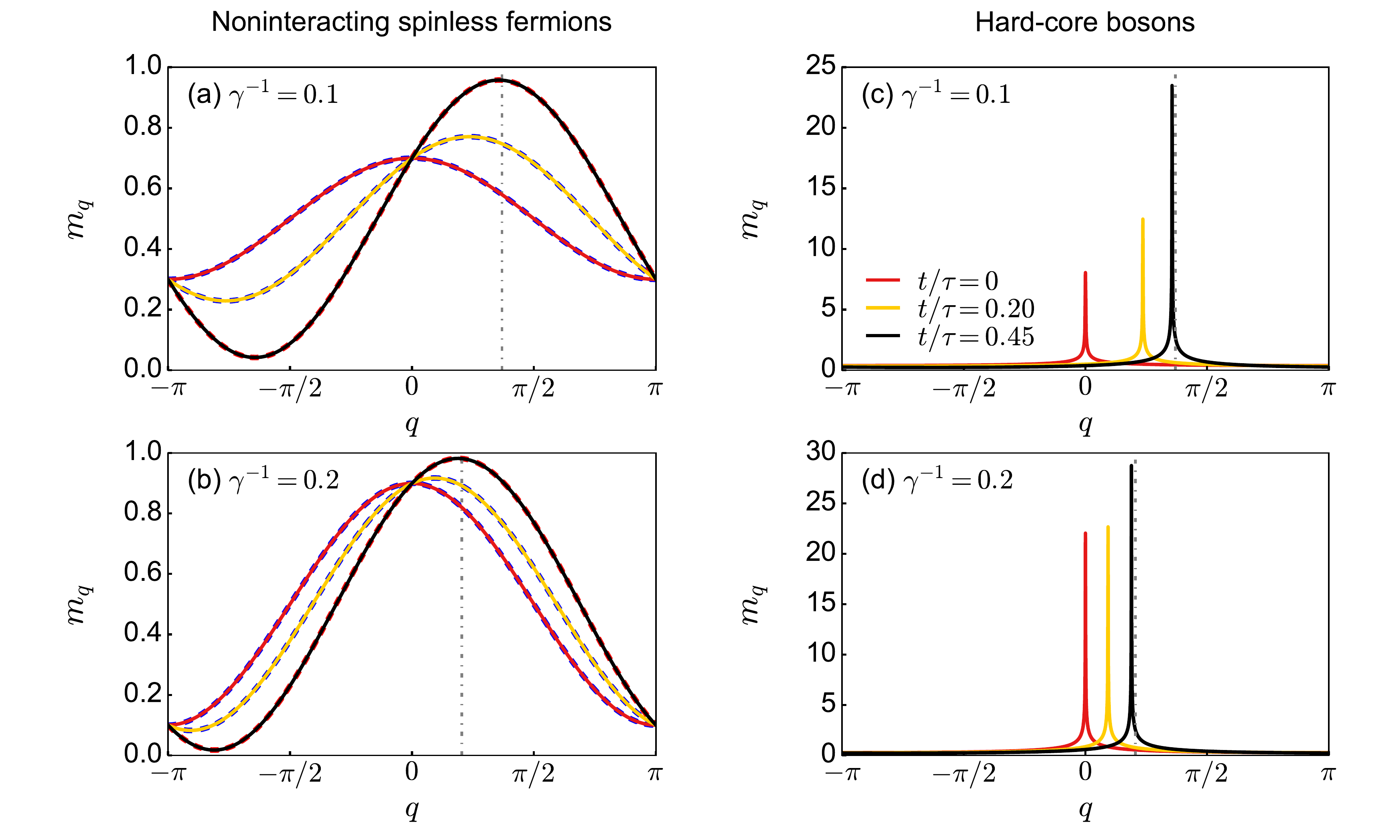}
  \vspace{-0.2cm}
  \caption{{\it Dynamics of the quasimomentum distribution function $m_q(t)$.}
(a) and (b): $m_q(t)$ for $N=1000$ noninteracting spinless fermions. (c) and (d): $m_q(t)$ for $N=1000$ hard-core bosons. Results are shown for $\gamma^{-1} = 0.10$ in (a) and (c), and for $\gamma^{-1} = 0.20$ in (b) and (d). Solid lines depict the numerical results, while the dashed lines in (a) and (b) were obtained using Eq.~\eqref{nq_free_t2}. Dotted vertical lines in all panels represent the shift of the peak in $m_q(t)$ at time $t/\tau=0.5$, and is given by Eq.~(\ref{def_phi_t_tau}).
   \label{fig3} }
  \end{center}
\end{figure*}

When computing the quasimomentum distribution function
\begin{equation} \label{def_mq}
 m_q(t) = \frac{1}{L} \sum_{j,l} e^{iq(j-l)}  \langle \psi(t)| {\cal \hat G}_{j,l} | \psi(t) \rangle,
\end{equation}
in which ${\cal \hat G}_{j,l} =\hat f_{j}^\dagger \hat f^{}_{l}$ for noninteracting fermions and ${\cal \hat G}_{j,l} = \hat b_{j}^\dagger \hat b^{}_{l}$ for hard-core bosons, not only do the absolute values of ${\cal \hat G}_{j,l}$ [see Fig.~\ref{fig2}] matter, but also their phases. Results for $m_q(t)$ are shown in Fig.~\ref{fig3} at different times and for different values of $\gamma$. The maximum of $m_q(t)$ exhibits two generic features during the dynamics: (i) it increases with time (coherence is dynamically enhanced), and (ii) its position shifts towards higher quasimomenta. 

In the ground state of the emergent local Hamiltonian~(\ref{Heme_free}), the peak position is determined by the phase $\varphi(t)$, which is identical for noninteracting fermions and hard-core bosons. We discuss its properties in Appendix~\ref{app1b} and Figs.~\ref{figsup1}(c)--\ref{figsup1}(d). Our analysis reveals that the peak emerges and stays at $q = \pi/2$ if the initial state is a sharp domain wall (a perfect product state, i.e., $\gamma^{-1}=0$). For all nonzero values of $\gamma^{-1}$, the position of the peak changes with time and is limited to quasimomenta below $\pi/2$.

For noninteracting fermions, increasing $\gamma^{-1}$ increases the spatial extent of the metallic interface in the initial state, which results in an increase of $m_{q=0}(t=0)$, see Figs.~\ref{fig3}(a) and~\ref{fig3}(b). In Appendix~\ref{app1c} we analytically show that
\begin{equation} \label{nq_free_initial}
 m_q(t=0) = \frac{N}{L} + 2 \gamma^{-1} \cos(q).
\end{equation}
The dynamics of the quasimomentum distribution $m_q(t)$ is obtained by a straightforward generalization of Eq.~(\ref{nq_free_initial}): (i) we change $q \to q(t) = q - \varphi(t)$ due to the time-dependent phase~(\ref{def_phase}), and (ii) replace $\gamma \to \gamma(t) = \gamma/{\cal A}(t) $ to account for the time-dependent hopping amplitude~(\ref{def_At}). This results in
\begin{equation} \label{nq_free_t2}
m_q(t) = m_q(t=0) + \frac{2t}{N} \sin(q).
\end{equation}
The predictions of Eq.~(\ref{nq_free_t2}), plotted as dashed lines in Figs.~\ref{fig3}(a) and~\ref{fig3}(b), are indistinguishable from the numerical results (solid lines) obtained by time-evolving the initial state.

The quasimomentum distribution $m_q(t)$ of hard-core bosons is markedly different from the fermionic one. This is a consequence of quasicondensation in the ground state $|\Psi_t \rangle$ of the emergent local Hamiltonian~(\ref{Heme_free}). Such a dynamical quasicondensation was studied theoretically in Ref.~\cite{rigol04}, in which the maximum value (in time) of the largest eigenvalue of the one-body density matrix was shown to scale as $\sqrt{N}$, and observed experimentally in Ref.~\cite{vidmar15}. The specific setup of Refs.~\cite{rigol04, vidmar15} is discussed in Sec.~\ref{sec_suddenexp}. 

Summarizing our presentation so far, we have introduced a physically relevant example in which the emergent eigenstate solution is applicable and for which the target eigenstate is the ground state of the emergent local Hamiltonian. This allowed us to understand why power-law correlations (and quasicondensation) can occur in transport problems far from equilibrium. It also helped us gain an analytic understanding of the behavior of quasimomentum distribution functions, for which phase factors in one-body correlations lead to peaks at nonzero quasimomenta.

\subsection{Heisenberg model} \label{sec_xxz}

The generality of the framework in Sec.~\ref{sec_formal} suggests that the emergent local Hamiltonian description is not restricted to noninteracting models or models mappable onto them. Here we demonstrate that this is indeed the case. We focus on one of the most widely studied models of quantum magnetism, the spin-1/2 XXZ chain. In contrast to the setup studied in Sec.~\ref{sec_free}, the (approximately) conserved operator $\hat Q$ in this case is not the particle current, but the energy current.

To make an explicit connection with the results obtained for noninteracting fermions, we map the spin-1/2 XXZ model onto interacting spinless fermions~\cite{cazalilla_citro_review_11}, and follow the notation from Sec.~\ref{sec_free}
\begin{eqnarray} \label{Hxxz}
\hat H_{V} & = & \sum_{l=-N+1}^{N-1} \hat h_l(V) \, ,\\
\hat h_l(V) & = & - \left(\hat f_{l+1}^\dagger \hat f^{}_l + {\rm H.c.} \right) + V \left( \hat n_l-\frac{1}{2}\right) \hspace*{-0.15cm} \left(\hat n_{l+1}-\frac{1}{2} \right), \nonumber
\end{eqnarray}
where $V$ denotes the amplitude of nearest-neighbor interaction. We consider systems with open boundaries and $L=2N$ lattice sites.

The operator $\hat B(V)$, which is the boost operator~\cite{zotos97} for the energy current $\hat Q(V)$, plays the role of $\hat P$ for noninteracting fermions. $\hat B(V)$ is defined as
\begin{equation}
\hat B(V) = \sum_{l=-N+1}^{N-1} l \, \hat h_l(V) \, ,
\end{equation}
and satisfies the relation $[\hat H_V, \hat B(V)] = i\hat Q(V)$. The energy current operator
\begin{eqnarray}\label{q3}
  \hat Q(V) &=& \sum_{l=-N+1}^{N-2} \left\{ \left(i \hat f_{l+2}^\dagger \hat f^{}_l + {\rm H.c.}\right)\right.\nonumber\\ &&- V \left(i \hat f_{l+1}^\dagger \hat f^{}_l + {\rm H.c.}\right) \left(\hat n_{l+2}-\frac{1}{2}\right)\nonumber\\&&  - \left. V \left(i \hat f_{l+2}^\dagger \hat f^{}_{l+1} + {\rm H.c.}\right) \left(\hat n_{l}-\frac{1}{2} \right) \right\}.\quad
\end{eqnarray}
is exactly conserved ($[\hat H_V,\hat Q(V)] = 0$) for periodic boundary conditions, and it is sometimes denoted as $\hat Q_3$ because each of its terms has support on three lattice sites~\cite{zotos97,mierzejewski15}.

\begin{figure*}[!t]
  \begin{center}
    \includegraphics[width=1.85\columnwidth]{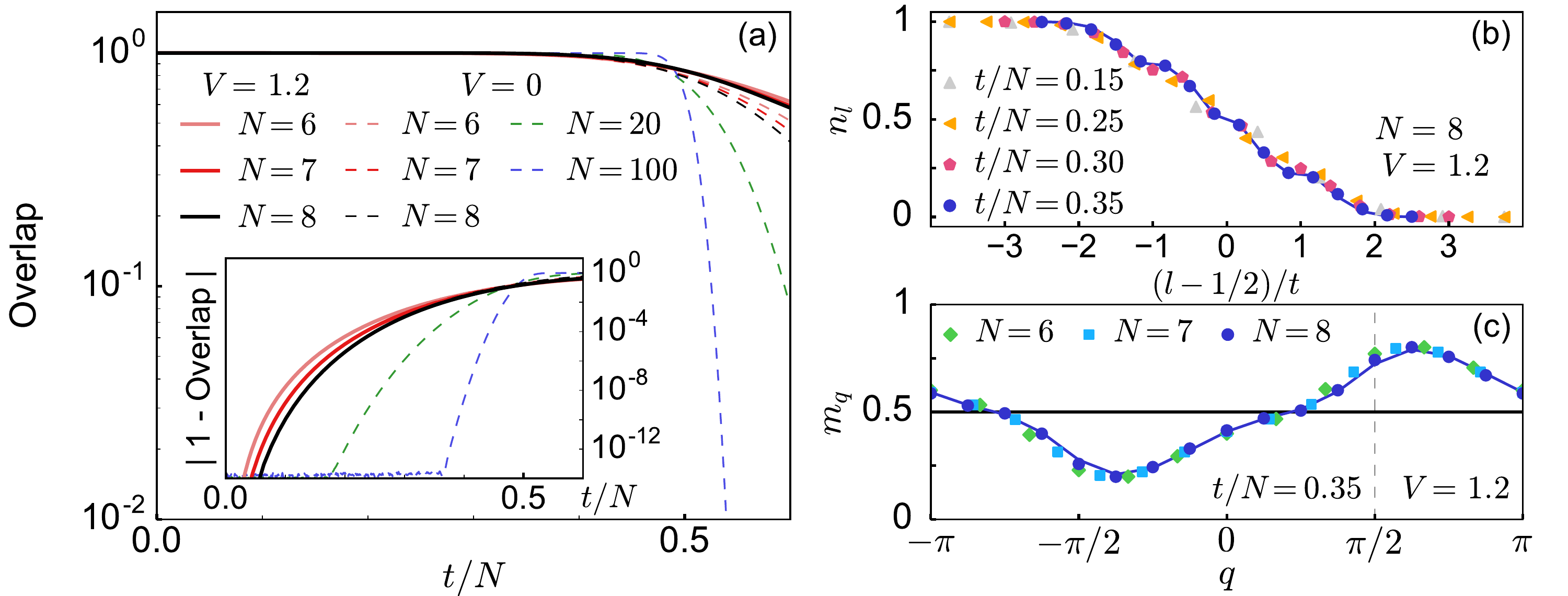}
  \caption{{\it Emergent Hamiltonian description of the domain-wall melting for interacting spinless fermions.} (a) Overlap $O(t) = |\langle \Psi^{V}_t | \psi^{V}(t) \rangle |$ as a function of rescaled time $t/N$ for $V=1.2$ (from full exact diagonalization) and $V=0$ (noninteracting system), and different system sizes. The inset depicts $|1-O(t)|$. For $N=6, 7,$ and $8$ (see inset), the results for interacting and noninteracting fermions are virtually indistinguishable at short times. Results for larger system sizes are only available for the noninteracting case. (b) and (c) Site and quasi-momentum occupations for $V=1.2$ obtained from $|\psi^{V}(t)\rangle$ (symbols) and from $|\Psi^{V}_t\rangle$ (lines; data shown for $N=8$ at time $t/N=0.35$). (b) Site occupations $n_l(t)$ for $N=8$ and different times as a function of the rescaled coordinate $(l-1/2)/t$. The data collapses onto a line~\cite{gobert05} as for noninteracting particles~\cite{vidmar15}. (c) Quasi-momentum distribution function $m_q(t)$ for different system sizes and the same rescaled time $t/N=0.35$.
  \label{fig4}}
   \end{center}
\end{figure*}

$\hat B(V)$ disconnects the left from the right part of the chain at the bond denoted by $l=0$. As a consequence, the sharp domain wall of spinless fermions, $|\psi_0\rangle = \prod_{l=-N+1}^0 \hat f_l^\dagger |\emptyset\rangle$, is an eigenstate of $\hat B(V)$ with eigenvalue $\lambda = 0$. We set the initial Hamiltonian $\hat H_0(V) = \hat B(V)$ and focus on the sharp domain wall as initial state so that Eq.~\eqref{def_initial} is satisfied with $\lambda=0$.

We construct the emergent local Hamiltonian ${\cal \hat H}_{V}(t)$ by following the derivation in Sec.~\ref{sec_formal}, starting from $\hat H_0(V)$. In analogy to Eq.~(\ref{Heme}), we define
\begin{equation} \label{Heme_xxz}
{\cal \hat H}_{V}(t) = \hat B(V) + t \, \hat Q(V) \, .
\end{equation}
We refer to the eigenstate of ${\cal \hat H}_{V}(t)$ that describes the dynamics as $| \Psi_t^V\rangle$.

The first question to be answered is the accuracy of the eigenstate $| \Psi_t^V\rangle$ to describe the time-evolving state $|\psi^V(t)\rangle$. We follow the analysis presented in Sec.~\ref{sec_almost}. In the setup considered here (i.e., a finite system with open boundaries), the operator $\hat Q(V)$ is not exactly conserved since
\begin{eqnarray} \label{xxz_order1}
\hspace*{-0.6cm}
[\hat H_V, \hat Q(V)] =  -i \bigg[  \left(1+ \frac{V^2}{4} \right) \bigg(\hat h_{-N+1}^{(1)} - \hat h_{N-1}^{(1)}\bigg) \nonumber \\
\hspace*{-0.6cm}  + V \bigg(\hat n_{-N+1} - \hat n_{-N+2}\bigg)^2 - V  \bigg(\hat n_{N-1} - \hat n_{N} \bigg)^2 \bigg],
\end{eqnarray} 
with $-N+1$ being the leftmost site. The expectation value of this commutator in the initial state~(\ref{Hheis_initial}) vanishes. To demonstrate the validity of the emergent eigenstate description for an extensive (in system size) time~(\ref{Hheis_almost1}), one needs to show that an extensive number of expectation values of higher-order commutators vanishes. This procedure is analogous to the one carried out for noninteracting fermions in Appendix~\ref{app1}. Even though obtaining a general expression for high-order commutators of the Heisenberg model is a daunting task, two consecutive commutators beyond Eq.~\eqref{xxz_order1} confirm that the largest support of the operators in Eq.~\eqref{Hheis_initial} grows linearly with the power of $t$ in Eq.~(\ref{Hheis_initial}), and that site occupation operators emerge in pairs (see Appendix~\ref{app2}). We therefore conjecture that the time regime in which the emergent Hamiltonian description is exponentially accurate increases with the size of the initial domain wall, and hence with the total particle number $N$, as for noninteracting fermions.

We complement the analysis above with numerical calculations. We use full exact diagonalization to calculate the time-evolved wavefunction $|\psi^V(t)\rangle$ and the eigenstate $|\Psi^V_t\rangle$ of the emergent local Hamiltonian $\hat {\cal H}_V(t)$. We add a small onsite potential to $\hat {\cal H}_V(t)$ in the leftmost site $(- 10^{-3}\, \hat n_{-N+1})$ to break the degeneracy of $|\Psi^V_t\rangle$ with other eigenstates with zero eigenenergy. This does not change $|\Psi^V_t\rangle$. It only changes its eigenenergy as $\langle \hat n_{-N+1} \rangle = 1$ (maximal site occupancy) for the times for which the emergent Hamiltonian description is valid.

In Fig.~\ref{fig4}(a), we plot the overlap between $|\Psi^{V}_t\rangle$ and $|\psi^{V}(t) \rangle$ for $V=0,\,1.2$ and different system sizes. All the overlaps are nearly one for $t/N\lesssim 0.5$ (see also the inset), independently of whether the system is interacting ($V=1.2$) or not ($V=0$). The results in the inset of Fig.~\ref{fig4}(a) reveal that, as $N$ increases, the ratio $t/N$ for which $|\Psi^{V}_t\rangle$ and $|\psi^{V}(t) \rangle$ are identical within computer precision increases. These results are qualitatively similar to those in Fig.~\ref{fig1}(a) and support the expectation that the emergent eigenstate description is also valid for interacting systems.

The emergent local Hamiltonian ${\cal \hat H}_V(t)$~(\ref{Heme_xxz}) includes the noninteracting point $V=0$. The sharp domain wall melting of noninteracting fermions can therefore be described either by an eigenstate $|\Psi_t\rangle$ of ${\cal \hat H}_{\rm SF}'(t)$~(\ref{Heme_free}) with $\gamma^{-1}= 0$, or by an eigenstate $|\Psi_t^{V=0}\rangle$ of ${\cal \hat H}_{V=0}(t)$. The two Hamiltonians describe different physics, nevertheless, they share at least one identical eigenstate $|\Psi_t\rangle = | \Psi_t^{V=0}\rangle$. Note, however, that $| \Psi_t\rangle$ used in Sec.~\ref{sec_free} is the ground state of ${\cal \hat H}_{\rm SF}'(t)$, while $| \Psi_t^V\rangle$ used here is a highly excited eigenstate of ${\cal \hat H}_V(t)$.

We now turn to physical properties of the current-carrying state. The shape of the propagating front has been already studied in the literature \cite{gobert05,sabetta_misguich_13,alba14}. In Fig.~\ref{fig4}(b), we plot the time evolution of the site occupations for $V=1.2$. The results collapse onto the same curve when the lattice positions are divided by time. For $V<2$~\cite{gobert05}, as in Fig.~\ref{fig4}(b), the site occupations do not differ significantly from the noninteracting case displayed in Fig.~\ref{fig2}(a). In contrast, the quasimomentum distribution [see Fig.~\ref{fig4}(c)] exhibits a pronounced difference with respect to the noninteracting case. This is because $m_q(t)$ develops a peak at $q>\pi/2$. For noninteracting fermions the peak emerges and remains at $q=\pi/2$ at all times. Further studies are needed to understand in a systematic way the effects of interactions and the role of initial states in long-lived current-carrying states of interacting spinless fermions.

\section{Sudden expansion of quantum gases in optical lattices} \label{sec_suddenexp}

In this section, we discuss the application of the emergent eigenstate description to the symmetric expansion of trapped ultracold spinless fermions, or hard-core bosons, after the harmonic confining potential is suddenly turned off but the optical lattice remains on. This is different from traditional time-of-flight measurements in which all potentials, including the optical lattice, are turned off and the particles expand in free space.

In recent experiments studying such setups, strongly-interacting bosons were initially prepared in one-dimensional Mott insulating states with one boson per site in the center of the trap~\cite{ronzheimer13,vidmar15}. For very strong on-site repulsive interactions, one can think of such bosons as hard-core bosons and their expansion dynamics is that of a domain wall ``melting'' symmetrically. So long as there remain sites with occupation one, the melting on each side can be described independently using the emergent eigenstate solution from Sec.~\ref{sec_free}. The question that remains is whether an emergent eigenstate solution exists after the occupation of the sites in the center of the system drops below one (but before the expanding fronts reach the edges of the lattice).

\begin{figure}
  \begin{center}
    \includegraphics[width=0.99\columnwidth]{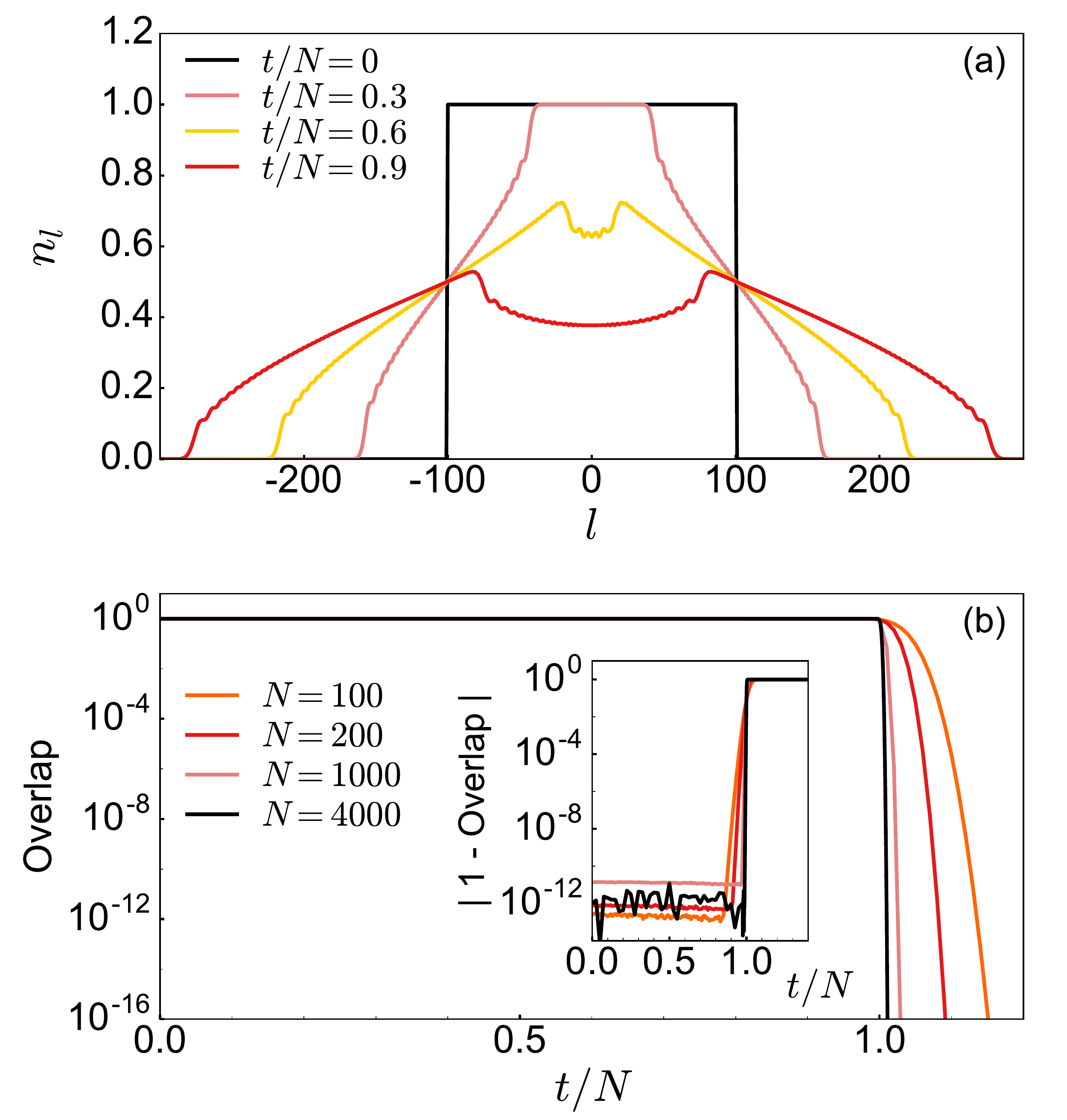}
    \end{center}
    \caption{ 
{\it Sudden expansion of noninteracting spinless fermions, or hard-core bosons, in optical lattices.} 
(a) Site occupations $n_l(t) = \langle \psi(t) | \hat n_l |\psi(t) \rangle$ at different times. (b) Overlap $O(t) = |\langle \Psi_t | \psi(t) \rangle |$ between the time-evolving state $|\psi(t)\rangle$ and the eigenstate $|\Psi_t \rangle$ of the emergent local Hamiltonian~(\ref{Heme_suddexp}), and $|1-O(t)|$ in the inset. The results are plotted versus the rescaled time $t/N$ for four system sizes. When increasing the number of particles $N$, $O(t)=1$ within machine precision (see inset) until $t/N \approx 1$.
    \label{fig5}}
\end{figure}

To answer that question, we study the dynamics of $2N + 1$ noninteracting spinless fermions (or, equivalently, hard-core bosons) on an open lattice with $L=6N+1$ sites, running from $-3N$ to $3N$. We consider the initial state $|\psi_0\rangle = \prod_{l\in L_0} \hat f_{l}^\dagger |\emptyset \rangle$, with the central sites $L_0 \equiv \{-N, -N+1, ..., N \}$ occupied, see Fig.~\ref{fig5}(a), and the remaining ones empty. In analogy to Sec.~\ref{sec_free}, the calculations can be straightforwardly extended to initial eigenstates that contain metallic (superfluid) domains surrounding the region with one fermion per site. Similar setups have been studied for hard-core bosons~\cite{rigol04, rigol05a, vidmar13}, as well as interacting soft-core bosons~\cite{rodriguez_manmana_06,vidmar13,brandino15} and spinful fermions~\cite{hm08,langer12,mei16}.

The initial product state $|\psi_0\rangle$ is an eigenstate of any Hamiltonian that is a sum of site occupation operators with arbitrary coefficients, which includes the linear and harmonic potentials. The emergent eigenstate solution can be constructed for both cases~\cite{vidmar_xu_17}. Here, we consider the linear potential, for which the analysis is simpler,
$\hat H_{0,\rm SE} = \hat P_{\rm SE}$, where the operator
\begin{equation}
\hat P_{\rm SE}= \frac{1}{L} \sum_{l=-3N}^{3N} l \, \hat n_l
\end{equation}
is essentially the one defined in Eq.~(\ref{def_Pfree}). An important difference w.r.t.~Sec.~\ref{sec_free} is that, here, the initial state is not the ground state of $\hat H_{0,\rm SE}$, but a highly excited state with eigenenergy $\lambda=0$. The dynamics is studied under
\begin{equation}
\hat H_{\rm SE}  = - \sum_{l=-3N}^{3N-1} (\hat f_{l+1}^\dagger \hat f^{}_l +{\rm H.c.}) \, .
\end{equation}
Figure~\ref{fig5}(a) shows the site occupations of the fermions, or, equivalently, hard-core bosons, at different times.

The emergent local Hamiltonian for this setup is
\begin{equation} \label{Heme_suddexp} 
{\cal \hat H}_{\rm SE}(t) = \frac{1}{L} \sum_{l=-3N}^{3N} l \, \hat n_l \, - \frac{t}{L} \sum_{l=-3N}^{3N-1} (i\hat f_{l+1}^\dagger \hat f^{}_l + {\rm H.c.} ) \, .
\end{equation}
Even though this Hamiltonian is essentially the one in Eq.~(\ref{Heme_free}) in the limit $\gamma^{-1}\to 0$, the target eigenstate $|\Psi_t \rangle$ lies in the center of the spectrum, and it is highly degenerate. Nevertheless, since the single-particle spectrum is nondegenerate, $|\Psi_t \rangle$ can be uniquely obtained as the Slater determinant comprising $2N+1$ consecutive single-particle states and giving $\lambda=0$. This is the way $|\psi_0 \rangle$ is constructed.

As for the domain wall melting, the emergent local Hamiltonian~(\ref{Heme_suddexp}) is in general not a conserved operator. The particle current operator does not commute with $\hat H_{\rm SE}$ due to the open boundary conditions~(\ref{HQfree}), which makes each higher-order commutator ${\cal \hat H}_{n,\rm SE}$ in Eq.~(\ref{def_Hheis_general}) nonzero. However, the support of operators in ${\cal \hat H}_{n,\rm SE}$ grows linearly from the lattice boundaries, as shown by the analysis in Appendix~\ref{app1}. As a result, the number of expectation values of ${\cal \hat H}_{n,\rm SE}$ that vanish in our initial state can be made arbitrarily large by just increasing the number of empty sites in the initial state. This results in the emergent eigenstate description being correct for arbitrarily long times independently of whether the site occupations in the center of the trap are one or below one during the expansion dynamics.

We verify the conclusions above by numerically calculating the overlap $O(t) = | \langle \Psi_t | \psi(t)\rangle |$ [see Fig.~\ref{fig5}(b)]. The results in Sec.~\ref{sec_free} for $\gamma^{-1}=0$ may lead one to conclude that the emergent eigenstate description will break down at $t/N \approx 0.5$ [see Fig.~\ref{fig1}(a)]. In contrast, in Fig.~\ref{fig5}(b), no change is observed in the dynamics of $O(t)$ at time $t/N = 0.5$, i.e., when the region with one atom per site in the center of the lattice has melted. Instead, $O(t)$ starts deviating from 1 around $t/N\approx1$. This is a consequence of particles reaching the lattice boundaries (the initial state has $2N$ empty sites on each side).

The emergent eigenstate description of the sudden symmetric expansion is therefore a nontrivial extension of the transport problems studied in Sec.~\ref{sec_free}. The fact that the dynamics for $t/N < 0.5$ within each half of the lattice in the symmetric expansion is identical to the domain-wall melting studied in Sec.~\ref{sec_free} allows us make an interesting observation. Namely, expectation values of observables in the ground state of Eq.~(\ref{Heme_free}) at $\gamma^{-1}\to 0$ are identical to the ones in a highly-excited eigenstate of Eq.~(\ref{Heme_suddexp}) for a system that is two times larger. As for the XXZ chain, this implies that power-law correlations, which in thermal equilibrium can only be observed in the ground state, also occur in highly-excited eigenstates of integrable Hamiltonians.

\section{Entanglement entropy of the target eigenstate} \label{sec_entanglement}

\begin{figure*}[!]
\begin{center}
    \includegraphics[width=1.7\columnwidth]{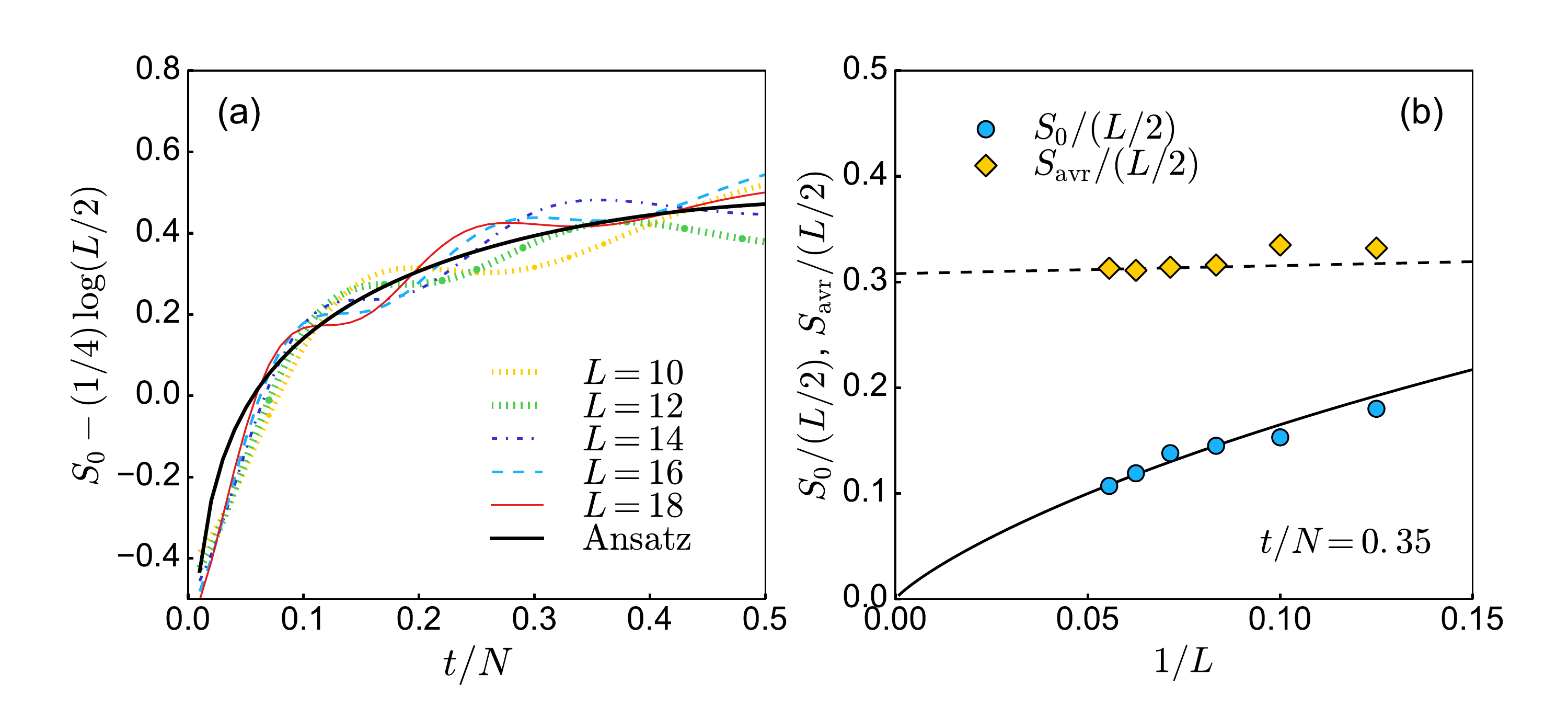}
    \vspace{-0.2cm}
\caption{{\it Entanglement entropy of eigenstates of the emergent local Hamiltonian $\hat {\cal H}_V(t)$~(\ref{Heme_xxz}) for $V = 1.2$.}
 (a) Entanglement entropy $S_0$ of the eigenstate $|\Psi_t\rangle$ that describes the domain-wall melting, after subtracting $(1/4) \log(L/2)$, for different system sizes $L$ and $N = L/2$. The thick solid line displays the ansatz~(\ref{See}) with $k' = 0.529$ as an average of $k'(L)$ for $L=14,16,18$. [For a given $L$, we obtain $k'(L)$ by fitting the numerical result to Eq.~(\ref{See}).] (b) Entanglement entropy per site at time $t/N = 0.35$ as a function of the inverse system size. Diamonds depict the average entanglement entropy $S_{\rm avr}$ per site of ${\cal D}/5$ eigenstates of the many-body spectrum (see text for details;  for $L=18$ we average over $10^3$ eigenstates). The solid line is obtained from Eq.~(\ref{See}) for the same value of $k'$ as in (a). The dashed line is a linear fit to $S_{\rm avr}/(L/2) = a_1/L + a_2$ for $L \geq 14$, with $a_1 = 0.076 $ and $a_2 = 0.308$.
\label{fig6} }
\end{center}
\end{figure*}

The evolution of the entanglement entropy in transport problems has been studied both for noninteracting~\cite{gobert05,eisler09, alba14} and interacting fermions~\cite{sabetta_misguich_13,alba14}. In the following, we focus on the domain-wall melting in the Heisenberg model and the underlying emergent local Hamiltonian ${\cal \hat H}_V(t)$~(\ref{Heme_xxz}). The entanglement entropy is defined as
\begin{equation}
 S = - {\rm Tr} \{ \hat \rho_{\rm L} \log{\hat \rho_{\rm L} } \}  = - {\rm Tr} \{\hat \rho_{\rm R} \log{\hat \rho_{\rm R}} \},
\end{equation}
where the reduced density matrix of the left/right subsystem is  $\hat \rho_{\rm L/R} = {\rm Tr}_{\rm R/L} \{ \hat \rho \}$, i.e., a trace over the complement subsystem of the total density matrix $\hat \rho$. The length of both subsystems is set to $L/2$. We focus on the entanglement entropy of excited eigenstates of the emergent local Hamiltonian $\hat {\cal H}_V(t)$, where the density matrix of eigenstate $n$ is $\hat \rho^{(n)} = |\Psi_t^{(n)} \rangle \langle \Psi_t^{(n)} |$. We are interested in the scaling of the entanglement entropy with the system size. In particular, we would like to distinguish between states that exhibit a volume-law scaling ($S/L = {\rm const}$  when $L \to \infty$), and states with vanishing entropy density ($S/L \to 0$ when $L \to \infty$), which include area-law states and critical states $S \sim \log(L)$.

We first focus on the entanglement entropy $S_0$ of the eigenstate of ${\cal \hat H}_V(t)$ that describes the domain-wall melting. Previous studies of the domain-wall melting in quantum spin chains suggested a logarithmic growth with time~\cite{sabetta_misguich_13,alba14}. Here, we are interested in the entanglement entropy at a fixed rescaled time $t/N$ to extract the dependence on the system size. Following Ref.~\cite{alba14}, we take the following ansatz
\begin{eqnarray} \label{See}
 S_0 & = & \frac{1}{4} \log{\left( \frac{L}{2} \right)} + \frac{1}{12} \log{\left( \frac{t}{N} \right)} \nonumber \\ 
 && + \; \frac{1}{6} \log{\left[ \sin\left( \frac{\pi t}{N}\right) \right]} + k',
\end{eqnarray}
where $k'$ is a constant. This heuristic ansatz is motivated by, but not equal to, the result for the local quench~\cite{eisler07,calabrese_cardy_07}. It was shown to provide accurate results, up to small temporal oscillations, for noninteracting fermions and the Heisenberg chain~\cite{alba14}.

Equation~(\ref{See}) predicts that, for a fixed $t/N$, the quantity $S_0 - (1/4) \log{\left( L/2 \right)} $ should be independent of the lattice size. The numerical results in Fig.~\ref{fig6}(a) are, up to small temporal fluctuations, consistent with this expectation. This confirms that the entanglement entropy per site $S_0/(L/2)$ vanishes in the thermodynamic limit [see circles in Fig.~\ref{fig6}(b)]. The novel aspect about this result is that it demonstrates that there are highly excited energy eigenstates of the interacting Hamiltonian ${\cal \hat H}_V(t)$ that exhibit a non-volume law scaling of the entanglement entropy. They are of relevance to transport problems like the ones studied here. Their contribution is expected to be negligible in the context of statistical mechanics.

On the other hand, the majority of eigenstates are expected to exhibit a volume-law scaling with system size. In Fig.~\ref{fig6}(b), we calculate the average entanglement entropy $S_{\rm avr}$ of ${\cal D}/5$ eigenstates of the many-body spectrum. Here, ${\cal D} = \binom{L}{N}$ is the Hilbert space dimension, and we choose the eigenstates $\{ |\Psi_t^{(m)}\rangle ; \, m \in [m_0 - {\cal D}/10, m_0 + {\cal D}/10] \}$, where $m_0$ is the eigenstate that describes the domain-wall melting. The dashed line in Fig.~\ref{fig6}(b) is a linear extrapolation of the entanglement entropy density vs $1/L$ to $L \to \infty$. It makes apparent that the entanglement entropy of the majority of the eigenstates of the emergent local Hamiltonian exhibits a volume-law scaling with the system size.

Our results thereby highlight the importance of gaining a better understanding of the entanglement entropy of highly-excited eigenstates of interacting integrable models~\cite{alba09, santos_polkovnikov_12, deutsch13, storms14, ares14, beugeling15, lai15,garrison15}. Eigenstates with non-volume-law scaling of the entanglement entropy exist throughout the spectrum, and they may play a special role in nonequilibrium states such as the ones studied here.

\section{Discussion and Outlook} \label{sec_conclusion}

The emergent eigenstate solution to quantum dynamics introduced in this work uncovers a new class of nonequilibrium states. The key property of this class of states is that they are eigenstates of a local operator we call the emergent Hamiltonian, which behaves as a conserved quantity. We have introduced a general framework that can be used to construct emergent local Hamiltonians. They are explicitly time dependent and do not commute with the physical Hamiltonian. 

We constructed emergent local Hamiltonians for simple experimentally relevant setups in which the Hamiltonian before the quench is either the boost operator or the boost operator plus the final Hamiltonian. Those setups are relevant to transport in systems with initial particle (or spin) imbalance, and to the sudden expansion of quantum gases in optical lattices. We should stress, however, that emergent local Hamiltonians can also be constructed to describe the dynamics of initial states that are stationary states of Hamiltonians that do not contain the boost operator~\cite{vidmar_xu_17}.

We used the emergent eigenstate solution in the context of noninteracting spinless fermions, hard-core bosons, and the Heisenberg model. We have shown that: (i) time-evolving current-carrying states can be ground states of emergent local Hamiltonians. The description of this family of states does not require one to invoke the concept of local equilibrium, which is often applied to describe nonequilibrium steady states. The quasimomentum distribution of those states exhibits a maximum at nonzero quasimomentum, which can be tuned by modifying the initial state. Our results explain the main features observed in a recent experiment with ultracold bosons in optical lattices~\cite{vidmar15}. (ii) Time-evolving states can also be highly-excited eigenstates of emergent local Hamiltonians, with entanglement entropies that do not exhibit volume law scaling. This highlights the physical relevance of some non-volume-law states in the bulk of the spectrum of integrable Hamiltonians to quantum dynamics.

There are problems of current interest in various fields to which the emergent eigenstate description introduced in this work can be applied: (i) The quench dynamics of thermal equilibrium states (relevant to theory and experiments dealing with ultracold quantum gases). Two recent studies of that problem were undertaken in Refs.~\cite{xu17,vidmar_xu_17}, in which (mixed) time-evolving states were shown to be Gibbs states of the emergent local Hamiltonian ${\cal \hat H}_{\rm SF}'(t)$ in Eq.~\eqref{Heme_free}, and an effective cooling observed numerically was explained analytically using ${\cal \hat H}_{\rm SF}'(t)$~\cite{xu17}. (ii) In the context of periodically driven systems, related ideas have been recently used to engineer integrable Floquet dynamics. Namely, to engineer driven systems that do not exhibit chaotic behavior under a periodic drive~\cite{gritsev17}. (iii) As pointed out by a referee, the construction introduced in this work can also be used to design fast forward Hamiltonians, which aim at bringing a system from the ground state of one Hamiltonian to the ground state of another Hamiltonian. This has potential applications in atomic, molecular and optical physics to design protocols that are much faster than the ones relying on adiabatic processes~\cite{Torrontegui13}. (iv) Even though the systems considered here are integrable, the emergent eigenstate construction only requires finding one conserved (or almost conserved) operator for a given initial state. This hints possible applications to (weakly) nonintegrable systems. The emergence of power-law correlations in the (nonintegrable) one-dimensional Bose-Hubbard model~\cite{rodriguez_manmana_06}, lends support to this expectation.

\begin{acknowledgments}
This work was supported by the Office of Naval Research. We acknowledge insightful discussions with M.~Mierzejewski, C.~D.~Batista, F.~Heidrich-Meisner, A. Polkovnikov, T.~Prosen, and U.~Schneider.
\end{acknowledgments}

\appendix

\section{Initial sharp domain wall} \label{app0}

A possible initial state to study current-carrying states is a sharp domain wall (a product state), in which the first $N$ sites ($L_0 \equiv \{-N, -N+1, \ldots, -1 \}$) of the lattice with open boundaries are occupied, and there are $N+1$ empty sites ($l\in\{0, 1, \ldots, N \}$). This state is an eigenstate of $\hat H_{0,\rm SF}$~(\ref{def_H0_sf}) in the limit $\gamma \to \infty$ in~(\ref{def_H0_sf}) and results in $\lambda = -(N+1)/2$. To avoid confusion with the limit $N \to \infty$, we consider in this case the initial Hamiltonian $\hat H_{0,\rm SF}'' = \hat P_{\rm SF}$ and the emergent local Hamiltonian
\begin{eqnarray}
{\cal \hat H}_{\rm SF}''(t) &=& \frac{1}{L} \sum_{l=-N}^{N} l \, \hat n_l - \frac{t}{L} \sum_{l=-N}^{N-1} (i \hat f_{l+1}^\dagger \hat f^{}_l + {\rm H.c.})\nonumber\\&&
 - \left( \lambda - \frac{t^2}{L} \right) \, .
\end{eqnarray}
To obtain the results for $\gamma^{-1}=0$, we therefore use the ground state $|\Psi_t \rangle$ of ${\cal \hat H}_{\rm SF}''(t)$ [relevant to Fig.~\ref{fig1}(a)] and calculate $\langle \psi(t) | {\cal \hat H}_{\rm SF}''(t) |\psi(t) \rangle$ [relevant to Fig.~\ref{fig1}(b)].

\section{Commutation relations for noninteracting spinless fermions} \label{app1}

The criterion~(\ref{Hheis_initial}) shows that the time regime in which the emergent Hamiltonian description~(\ref{Heme_free}) can be applied is related to the expectation values of higher-order commutators ${\cal\hat H}_{n,\rm SF}$ in the initial state. We consider initial states that are ground states of $\hat H_{0,\rm SF}$~(\ref{def_H0_sf}) at $\gamma > \gamma^*$, i.e., they exhibit extensive (in system size) regions of site occupancy one/zero at the lattice boundaries. To understand the structure of ${\cal\hat H}_{n,\rm SF}$, it is useful to evaluate a few terms beyond ${\cal \hat H}_{1,\rm SF}$~(\ref{HQfree}). The operator ${\cal \hat H}_{2,\rm SF}$ is proportional to
\begin{equation} 
{\cal \hat H}_{2,\rm SF} \propto [\hat H_{\rm SF}, (\hat n_{-N} - \hat n_{N})] = i \left( \hat \jmath_{-N}^{(1)} + \hat \jmath_{N-1}^{(1)} \right), \label{t3}
\end{equation}
where we have defined a generalized current operator with support on $m+1$ sites as: $\hat \jmath_l^{(m)} = (i\hat f_{l+m}^\dagger \hat f^{}_l + {\rm H.c.}$). Analogously, one can define a generalized kinetic energy operator: $\hat h_l^{(m)} = ( \hat f_{l+m}^\dagger \hat f^{}_l + {\rm H.c.})$.  In Eq.~\eqref{t3}, the current operators have support on two sites at the chain boundaries. Hence, the expectation values in the initial state yield zero.

New terms in higher-order commutators are generated symmetrically at both ends of the chain, so let us focus on the left end. The operator ${\cal \hat H}_{3,\rm SF}$ contains
\begin{equation} 
[\hat H_{\rm SF}, \hat \jmath_{-N}^{(1)}] = -i \left[ \hat h_{-N}^{(2)} + 2(\hat n_{-N} - \hat n_{-N+1}) \right].
\end{equation}
Again, this term vanishes in the initial state.

Moving forward, the following commutation relations are useful:
\begin{eqnarray} \label{hjm}
  &&[\hat H_{\rm SF}, \hat \jmath_{l}^{(m)}] = -i \times\\ &&\qquad
  \left[ \left(\hat h_{l}^{(m+1)} - \hat h_{l-1}^{(m+1)}\right) +
    \left(\hat h_{l}^{(m-1)} - \hat h_{l+1}^{(m-1)} \right) \right],
    \nonumber
\end{eqnarray}
for the generalized current operator,
\begin{eqnarray}  \label{hhm}
  &&[\hat H_{\rm SF}, \hat h_{l}^{(m)}] = i\times\\ &&\qquad 
  \left[ \left(\hat \jmath_{l}^{(m+1)} - 
  \hat \jmath_{l-1}^{(m+1)} \right) + \left(\hat \jmath_{l}^{(m-1)} - 
  \hat \jmath_{l+1}^{(m-1)} \right) \right], 
  \nonumber
\end{eqnarray}
for the generalized kinetic energy operator ($m>0$), and
\begin{equation}
  \label{hn} [\hat H_{\rm SF}, \hat n_{l}] =i \left( \hat \jmath_{l}^{(1)} - \hat \jmath_{l-1}^{(1)} \right) ,
\end{equation}
for the site occupation operator. In Eqs.~\eqref{hjm}--\eqref{hn}, the operator at site $l-1$ vanishes if $l=-N$.

One can see from Eqs.~\eqref{hjm}--\eqref{hn} that the maximal support of the operators in ${\cal \hat H}_{n,\rm SF}$, which are all one-body operators, grows linearly with $n$. Moreover, from Eq.~\eqref{hjm} for $m=1$, one can see that site occupation operators $\hat n_{l}$ and $\hat n_{l+1}$ emerge in pairs. Since the initial states under consideration are product states at the boundaries, this advances that an extensive (in system size) number of terms in Eq.~(\ref{Hheis_initial}) vanishes.

For concreteness, let us focus on initial states that are sharp domain walls (as defined in Appendix~\ref{app0}). The site occupation operators $\hat n_{l}$ and $\hat n_{l+1}$, which emerge in pairs, cancel each other in sharp domain walls unless $l=-1$. This occurs for the first time in operator ${\cal \hat H}_{2N+1,\rm SF}$. Hence, the first nonvanishing expectation value in Eq.~(\ref{Hheis_initial}) (after $t^2$) is $\langle \psi_0| {\cal \hat H}_{2N+1,\rm SF}|\psi_0\rangle$. The coefficient in front of this term is $t^{2N+2}/(2N+2)!$, and for the validity of the emergent eigenstate description it is sufficient that it be exponentially small. Using Stirling's formula ($N\gg1$), we get the condition $t \lesssim N(2/e)$. By comparison to exact numerical results in Fig.~\ref{fig1}(a), we show that this analysis correctly predicts the time of validity of the emergent eigenstate description to be proportional to the particle number $N$ (or the system size $L$).

\section{Quasimomentum distribution of noninteracting fermions in a linear potential} \label{app1c}

Here, we derive Eq.~(\ref{nq_free_initial}). We study the Wannier-Stark Hamiltonian in the infinite volume limit,
\begin{equation}
  \label{eq:1}
  \hat H = -\sum_{l = -\infty}^{\infty}\left(\hat f^{\dag}_{l}\hat f_{l+1} + \text{h.c.}\right) +\frac{\gamma}{L'} 
  \sum_{-\infty}^{\infty} l \, \hat n_{l} \, ,
\end{equation}
where $\gamma$ is finite, and $L'/\gamma$ determines the width of the localized states. These localized single-particle Wannier-Stark eigenstates can be written as
\begin{equation}
  \hat \eta_n^\dagger |\emptyset \rangle = \sum_{l=-\infty}^{\infty} J_{l-n}(\alpha) \hat f_l^\dagger |\emptyset \rangle \, ,
\end{equation}
where $n$ is a site-index, $J_{l-n}(\alpha)$ is the Bessel function of the first kind, $|\emptyset \rangle$ is the vacuum state, and we define $\alpha \equiv 2L'/\gamma$. The eigenstate above represents a wave function localized at site $n$ with a width given by $\alpha$. In order to produce the domain-wall initial state, we consider a many-body state $|\psi_0'\rangle = \prod_{n=-\infty}^{-1} \hat \eta_n^\dagger |\emptyset \rangle$ with $2\alpha < L'$, i.e., $\gamma > 4$.

One-body correlations $C_{l,m} = \langle \psi_0' | \hat f_l^\dagger \hat f_m | \psi_0' \rangle$ for $l\neq m$ can be expressed as
\begin{eqnarray} \label{eq:3} C_{l,m} & = &
  \sum_{n=-\infty}^{-1}
  J_{l-n}(\alpha)J_{m-n}(\alpha)\\ &=&
  \frac{\alpha}{2(l-m)}\left[J_{m+1}(\alpha)J_{l}(\alpha) -
    J_{m}(\alpha)J_{l+1}(\alpha) \right] \, .\nonumber \label{clm_exact}
\end{eqnarray}
for any $l, m\in\mathbb{Z}$ with $l\neq m$. Note that $C_{l,m}\approx 0$ for $|l|,|m| \gg \alpha$. We use $C_{l,m}$ to calculate the contribution of the off-diagonal matrix elements of the one-body density matrix to $m_q$, which we define as
\begin{equation}
m_q = \lim_{N'\to\infty}\frac{1}{L'}\sum_{l,m=-N'}^{N'} e^{iq(l-m)}C_{l,m} \, .
\end{equation}
This formula calculates the quasimomentum distribution on a line of length $L'=2N'+1$. In the limit $N'\to\infty$, $L'$, which also defines the width of the localized Wannier-Stark states, goes to infinity. This is required in order to correctly define the thermodynamic limit of this system. If the width of the Wannier-Stark states, $\alpha$, did not grow as $L'$, the state would go to a step-like domain-wall in the $N'\to\infty$ limit and render the quasimomentum distribution trivial.

\begin{figure*}[!t]
\begin{center}
\includegraphics[width=1.7\columnwidth]{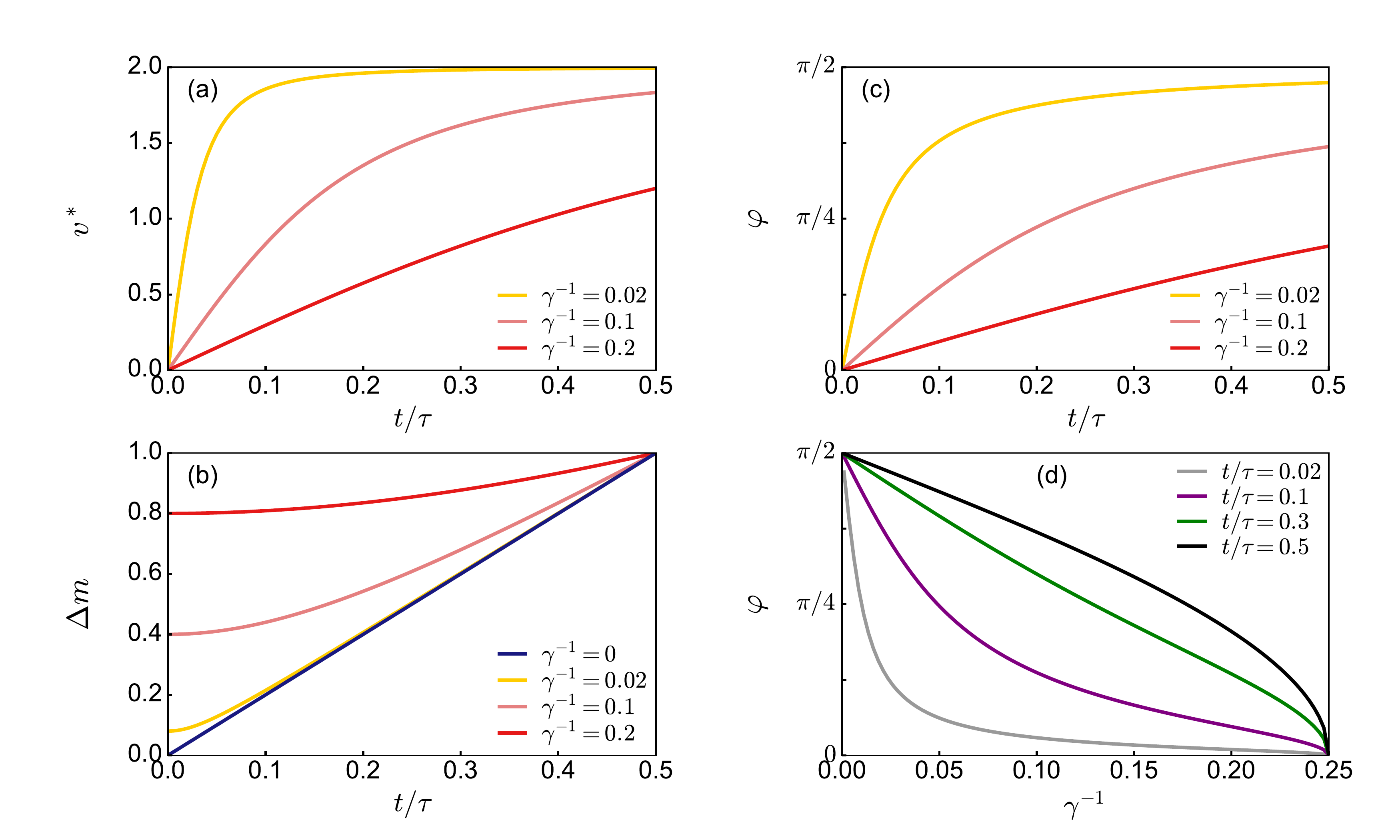}
\vspace{-0.2cm}
\caption{{\it Site and quasimomentum occupations of noninteracting spinless fermions for $N/L=1/2$.}
(a): Velocity of the propagating front $v^*(t)$~(\ref{def_vstar}). (b): Amplitude of the quasimomentum distribution function for noninteracting fermions $\Delta m(t)$~(\ref{def_delta_nq}). (c) and (d): Position of the peak of the quasimomentum distribution function $\varphi(t)$~(\ref{def_phi_t_tau}). Results are shown versus $t/\tau$ in (a)--(c), and versus $\gamma^{-1}$ in (d).
\label{figsup1}}
\end{center}
\end{figure*}

The diagonal contribution to $m_q$ is given by the average site occupation which for the half-filled case considered here is always \nicefrac12. We define the off-diagonal contribution as $\bar m_q \equiv m_q - 1/2$, so that
\begin{eqnarray}
  \label{eq:4}
    \bar m_q &=& \lim_{N'\to\infty}\frac{1}{L'}\sum_{l\neq m=-N'}^{N'} e^{iq(l-m)} C_{l,m} \nonumber\\ &=& \lim_{N'\to\infty}
    \frac{1}{L'}\sum_{l\neq m=-N'}^{N'} \frac{\alpha \, e^{iq(l-m)}}{2(l-m)}\\&& \hspace{1cm}\times\left[J_{m+1}(\alpha)J_{l}(\alpha) - J_{m}(\alpha)J_{l+1}(\alpha) \right] \, .\nonumber
\end{eqnarray}
We express $1/(l-m)$ (for $l\neq m$) as an integral, $1/(l-m) = \int_{0}^{1}{\rm d}x\, x^{l-m-1}$, giving
\begin{widetext}
\begin{equation}
  \label{eq:8}
  \begin{split}
    \bar m_{q} &= \lim_{N'\to\infty}\frac{\alpha}{2L'}
    \int_{0}^{1}  {\rm d}x \sum_{l, m=-N'}^{N'} e^{iq(l-m)}x^{l-m-1} \left[J_{m+1}(\alpha)J_{l}(\alpha) - J_{m}(\alpha)J_{l+1}(\alpha) \right] \\
    &= \lim_{N'\to\infty}\frac{\alpha}{2L'} \int_{0}^{1} {\rm d}x
    \sum_{l, m=-N'}^{N'}
    \big[e^{iq}(xe^{iq})^{l}(xe^{iq})^{-(m+1)}J_{m+1}(\alpha)J_{l}(\alpha) \\ 
      &\hspace{5.7cm}-
      x^{-2}e^{-iq}(xe^{iq})^{l+1}(xe^{iq})^{-m}J_{m}(\alpha)J_{l+1}(\alpha)
    \big]
  \end{split}
\end{equation}
\end{widetext}
Using the generating function for Bessel functions of the first kind, $\sum_{n=-\infty}^{\infty}J_{n}(\alpha)x^{n} = \exp[\alpha(x-x^{-1})/2]$, reinstating $\alpha=2L'/\gamma$, and taking the limit, we get,
\begin{equation}
  \label{eq:9}
  \begin{split}
    \bar m_{q} &= \gamma^{-1}    \int_{0}^{1}  {\rm d}x \left(e^{iq} - x^{-2}e^{-iq} \right) \\
    &= 2\gamma^{-1}\cos q \, .
  \end{split}
\end{equation}
As a result, the quasimomentum distribution function reads
\begin{equation}
  m_q = \frac{1}{2} + 2\gamma^{-1} \cos q \, .
\end{equation}

\section{Site and quasimomentum occupations for noninteracting fermions} \label{app1b}

Here we provide further details, which complement the analysis in Sec.~\ref{sec_free}, about the time evolution of the site and quasimomentum occupations in the current-carrying states.

We first derive the time $t^*$ at which the site occupations at the boundaries of the chain depart from their initial values. In the ground state of ${\cal \hat H}_{\rm SF}'(t)$, $t^*$ can be obtained from the Wannier-Stark solution using that $\gamma/{\cal A}(t^*) = \gamma^*$ ($\gamma^* = 4$ for $L/N=2$). Substituting ${\cal A}(t^*)$ from Eq.~(\ref{def_At}), this expression can be simplified to $t^*/L = (\gamma^*)^{-1} \sqrt{1 - (\gamma^*/\gamma)^{2}}$. In the case of $\gamma^{-1}=0$, it yields $t^*/L=1/\gamma^*$. By comparing that to  $t^*/N = 1/v_{\rm max}=1/2$, one can see that $\gamma^*/v_{\rm max}=L/N$. We choose the unit of time $\tau$ such that $t^*$ satisfies $t^*/\tau := 1/v_{\rm max}$. This results in $\tau = N (t^*/L) \gamma^*$, or Eq.~(\ref{def_tau}) for the particular case in which $L/N=2$.

The site occupations shown in Fig.~\ref{fig2}(a) exhibit data collapse upon replacing the site positions $l \to \tilde l = l \gamma(t)/(2L)$, where $\gamma(t) = \gamma/{\cal A}(t)$. This scaling enables one to obtain a closed expression for the position $l^*(t)$ of the propagating front by setting $l^*(t) \gamma(t)/(2L) = 1$. It also allows one to determine the velocity of the propagating front, $v^*(t) = {\rm d}l^*(t)/{\rm d}t$, which can be expressed in units of $t/\tau$ as
\begin{equation} \label{def_vstar}
 v^*(t) = 2 \; \frac{\frac{t}{\tau} \frac{N}{L} \sqrt{1- \left( \gamma^*\gamma^{-1} \right)^2}}{\sqrt{\gamma^{-2} + \left( \frac{t}{\tau} \right)^2 \left( \frac{N}{L} \right)^2 \left[ 1- \left( \gamma^*\gamma^{-1} \right)^2 \right]}} \; .
\end{equation}
This expression makes clear that $v^*(t) = 2$ only when $\gamma^{-1}=0$, i.e., for the initial product state. On the other hand, for nonzero $\gamma^{-1}$, one gets that $v^*(t) < 2$. The time evolution of $v^*$ is plotted in Fig.~\ref{figsup1}(a) for three values of $\gamma$.

Next, we study the quasimomentum distribution $m_q(t)$~(\ref{def_mq}). We focus on two quantities: (i) the time evolution of $\varphi(t)$~(\ref{def_phase}), which gives the position of the peak in $m_q(t)$, and (ii) the amplitude 
\begin{equation} \label{def_delta_nq}
\Delta m(t) = \mbox{Max}[m_q(t)] - \mbox{Min}[m_q(t)]
\end{equation}
of $m_q(t)$ for noninteracting spinless fermions.

The time-dependent renormalization of the amplitude of $m_q(t)$, invoked in the derivation of Eq.~(\ref{nq_free_t2}), suggests that $\Delta m(t) = 4 {\cal A}(t)/\gamma = 2 \sqrt{\gamma^{-2} + (t/L)^2}$. Hence, $\Delta m(t)$ increases linearly with $t$ only when $\gamma^{-1}=0$ (the initial amplitude is zero), yielding $\Delta m(t) = 4t/L$. In all other cases, the amplitude increase is slower than linear. 

By expressing $\Delta m$ in units of $t/\tau$, one gets
\begin{equation}
 \Delta m(t) = 4 \sqrt{\gamma^{-2} + \left( \frac{t}{\tau} \right)^2 \left( \frac{N}{L} \right)^2 \left[ 1 - \left( \gamma^* \gamma^{-1} \right)^2 \right]} \; ,
\end{equation}
which is plotted in Fig.~\ref{figsup1}(b). Note that $\Delta m(t) = 1$ for $t^*/\tau=0.5$, implying that, at the time at which the expanding front reaches the chain boundary, there exist one quasimomentum that is fully occupied and one that is empty.

The position of the peak of the quasimomentum distribution in the ground state of ${\cal \hat H}_{\rm SF}'(t)$~(\ref{Heme_free}) is determined by the phase $\varphi(t)=\arctan(\gamma t/L)$. This result reveals that the position of the peak during the expansion of a domain wall evolves from $q=0$ towards $q=\pi/2$, with the only exception being the initial product state, $\gamma^{-1}=0$. In that case, the peak is located at $q=\pi/2$ at all times $t>0$. In terms of $t/\tau$, one can write the phase $\varphi$ as
\begin{equation} \label{def_phi_t_tau}
 \varphi(t) = \arctan\left[ \frac{t}{\tau} \frac{N}{L} \sqrt{\gamma^2 - (\gamma^*)^2} \right] \; .
\end{equation}
We plot $\varphi$ versus $t/\tau$ in Fig.~\ref{figsup1}(c), and versus $\gamma^{-1}$ in Fig.~\ref{figsup1}(d). The plots show that, for any nonzero value of $\gamma^{-1}$, the position of the peak in $m_q(t)$ remains below $q=\pi/2$ at all times.

\section{Commutation relations in the Heisenberg model} \label{app2}

In this Appendix, we provide explicit expressions for two consecutive nested commutators of the operator $\hat Q(V)$~(\ref{q3}) with $\hat H_V$~(\ref{Hxxz}). The result for the second-order commutator on the left boundary of the chain is given by:
\begin{eqnarray}
 &&\bigl[\hat H_V, \bigl[\hat H_V, \hat Q(V) \big] \big] \bigg|_{\rm left \; boundary} =  \hat \jmath_{-N+1}^{(2)} \nonumber\\
 && - V \left[ \hat \jmath_{-N+1}^{(1)}  \left( \hat n_{-N+3} - \frac{1}{2} \right) + 2 \hat \jmath_{-N+2}^{(1)} \left( \hat n_{-N+1} - \frac{1}{2} \right) \right]\nonumber \\
 && + \frac{V^2}{4} \hat \jmath_{-N+1}^{(2)} - \frac{V^3}{4} \hat \jmath_{-N+1}^{(1)} \left( \hat n_{-N+3} - \frac{1}{2} \right).
\end{eqnarray}
In the latter expression, the maximal support of the operators extends on the three leftmost lattice sites, and gives zero when acting on the initial state. In the same manner, the third-order commutator acts at most on the four leftmost lattice sites,
\begin{widetext}
\begin{equation} \label{xxz_order3}
  \begin{split}
    & \bigl[ \hat H_V, \bigl[\hat H_V, \bigl[\hat H_V, \hat Q(V) \big] \big] \big] \bigg|_{\rm left \; boundary} = i \Biggl\{ - \left[ \hat h_{-N+1}^{(3)} + (\hat h_{-N+1}^{(1)} - \hat h_{-N+2}^{(1)}) \right]\\
    & + V \bigg[ \hat h_{-N+1}^{(2)} \left( \hat n_{-N+4}-\frac{1}{2} \right) + 2 \hat h_{-N+2}^{(2)} \left(\hat n_{-N+1} - \frac{1}{2}\right) - \hat h_{-N+1}^{(2)} \left(\hat n_{-N+2} - \frac{1}{2} \right) - \hat \jmath_{-N+1}^{(1)} \hat \jmath_{-N+3}^{(1)} \\
    & \hspace*{1.0cm} + 2 (\hat n_{-N+1} - \hat n_{-N+2}) \left( \hat n_{-N+3} - \frac{1}{2} \right) + 2(\hat n_{-N+2} - \hat n_{-N+3}) \bigg] \\
    &  - \frac{V^2}{4}  \left[  \hat h_{-N+1}^{(3)} + (\hat h_{-N+1}^{(2)} - \hat h_{-N+2}^{(1)}) + 4 \hat h_{-N+1}^{(1)} \left( \hat n_{-N+3} - \frac{1}{2} \right) + 8 \hat h_{-N+2}^{(1)} \left( \hat n_{-N+1} - \frac{1}{2} \right) \left( \hat n_{-N+4} -\frac{1}{2} \right) - 2\hat h_{-N+2}^{(1)} \right]  \\
    & + \frac{V^3}{4}  \left[  \hat h_{-N+1}^{(2)} \left( \hat n_{-N+4} - \frac{1}{2} \right) + \hat h_{-N+1}^{(2)} \left( \hat n_{-N+2} - \frac{1}{2} \right) -  \hat \jmath_{-N+1}^{(1)} \hat \jmath_{-N+3}^{(1)} + 2 (\hat n_{-N+1} - \hat n_{-N+2}) \left( \hat n_{-N+3} - \frac{1}{2} \right)  \right] \\
    & - \frac{V^4}{4} \hat h_{-N+1}^{(1)} \left( \hat n_{-N+3} - \frac{1}{2} \right) \Biggr\},
  \end{split}
\end{equation}
\end{widetext}
giving again zero when acting on the initial state. Locality of the Hamiltonian $\hat H_V$~(\ref{Hxxz}) guaranties that in every new generation, the support of operators will increase at most for one site, similarly to noninteracting fermions in Appendix~\ref{app1}.  As for the noninteracting case, we then conjecture the times for which the emergent Hamiltonian description of the Heisenberg model is exponentially accurate to increase nearly linearly with $N$. The numerical results in Fig.~\ref{fig4}(a) confirm this hypothesis.

\bibliographystyle{biblev1}
\bibliography{references}

\end{document}